\documentclass[twocolumn]{aastex61}

\newcommand\aastex{AAS\TeX}

\newcommand{\eqref}[1]{ (\ref{#1}) }
\def\Ra{\mbox{\rm Ra}}

\def\urms{u_{\rm rms}}

\def\nueff{\nu_{\rm eff}}
\def\nut{\nu_{\rm t}}
\def\Re{{\rm Re}}

\def\apjl{ApJL}
\def\apjs{ApJ}
\def\aap{A\&A}

\def\Rs{R_{\odot}}

\newcommand{\brac}[1]{\langle #1 \rangle}

\usepackage{natbib}
\usepackage{xcolor}
\usepackage{bm}
\usepackage{enumerate}
\usepackage{hyperref}
\usepackage{float}
\usepackage[utf8]{inputenc}
\usepackage{graphicx}
\usepackage{graphics}
\graphicspath{{eps/}{png/}}

\shorttitle{\aastex\ Convergence in ILES of convection}
\shortauthors{Nogueira et al.}
\begin{document}

\title{Numerical convergence of 2D solar convection in implicit large-eddy simulations}

\correspondingauthor{G. Guerrero}
\email{guerrero@fisica.ufmg.br}  

\author{H.~D. Nogueira}
\affiliation{Physics Department, Universidade Federal de Minas Gerais \\
Av. Antonio Carlos, 6627, Belo Horizonte, MG 31270-901, Brazil}
\affiliation{Laboratoire Kastler Brossel, Sorbonne Universit\'e, CNRS, ENS-Universit\'e PSL, \\
 Coll\`ege de France, 4 place Jussieu, F-75005 Paris, France}   

\author{G. Guerrero}
\affiliation{Physics Department, Universidade Federal de Minas Gerais \\
Av. Antonio Carlos, 6627, Belo Horizonte, MG 31270-901, Brazil}
\affiliation{New Jersey Institute of technology, Newark, NJ 07103,USA}

\author{P.~K. Smolarkiewicz}
\affiliation{National Center for Atmospheric Research, Boulder, Colorado, USA}   

\author{A.~G. Kosovichev}
\affiliation{New Jersey Institute of technology, Newark, NJ 07103,USA}

\begin{abstract}
Large-eddy simulations (LES) and implicit LES (ILES) are wise and affordable alternatives to 
the unfeasible direct numerical simulations (DNS) of turbulent flows at high Reynolds numbers
(Re).  However, for systems with few observational constraints, it is a formidable challenge 
to determine if these strategies adequately capture the physics of the system. Here we address 
this problem by analyzing numerical convergence
of ILES of turbulent convection in 2D, with resolutions between $64^2$ and $2048^2$ grid
points,  along with the estimation of their effective viscosities,
resulting in effective Reynolds numbers between $1$ and $\sim10^4$. 
The thermodynamic structure of our model resembles the 
solar interior, including a fraction of the radiative zone and the convection zone. 
In the convective layer, the ILES solutions converge for the simulations with $\ge 512^2$ 
grid points, as evidenced by the integral properties of the flow and its power spectra. 
Most importantly, we found that even a resolution of $128^2$ grid points, $\Re\sim10$, is 
sufficient to capture the dynamics of the large scales accurately. This is a consequence
of the ILES method allowing that the energy contained in these scales
is the same in simulations with low and high resolution.
Special attention is needed in regions with a small density scale height driving the 
formation of fine structures unresolved by the numerical grid. In the stable layer
we found the excitation of internal gravity waves, yet high 
resolution is needed to capture their development and interaction.  
\end{abstract}


\section{Introduction} 
\label{sec:intro}

Turbulent convection is ubiquitous in astrophysics and geophysics, taking place in 
planetary atmospheres and oceans, and inside stars. It is a non-linear problem for 
which analytic solutions are scarce and limited in scope.
On the other hand, laboratory experiments are able to explore only a fraction of the parameter 
space spanned by convection in nature.
Thus, many scientists turn to computer simulations to investigate this phenomenon. 
The simulations are carried out for ideal situations, such as Rayleigh-Bénard 
convection \citep{stevens2018turbulent, sakievich2016large}, as well as for several cases 
observed in nature ranging from stellar convection 
\citep[e.g.,][]{2002IJNMF..39..855E, 2005AN....326..681B,
2016ApJ...818...32F, 2016ApJ...819..104G, 2016ApJ...828L...3G, 2016ApJ...821L..17K,
kapyla+19} to 
the Earth's and planetary atmospheres as well 
as oceans and mantle currents \citep{2014AgFM..184...12G, 2019JCli...32.7453P, 2019JMetR..33..949W}.

The simulations resolving all the relevant scales of the flow are called  
direct numerical simulation (DNS). In DNS, the resolved scales range from the largest scale 
to the smallest  Komolgorov scale, at which dissipation occurs. In high Reynolds number systems, 
such as in the  Solar convection zone,  the ratio between the largest and the smallest 
scales is extremely large. Consequently, to capture all relevant scales, the domain
discretization must be excessively fine \citep[$N^D \sim \Re^{3D/4}$, where 
$N$ is the number of grid points, and $D$ is the number of spatial dimensions,][]{2000tufl.book.....P}.
These simulations are still unfeasible for modern supercomputers.
The expectation is that with progressively increasing the resolution 
the prognostic variables reach an asymptotic regime of convergent values for $N^D \ll \Re^{3D/4}$. 
This would mean that the relevant scales for a given system are larger than
the Kolmogorov scale, and that the dissipative processes are governed by
turbulence. For instance, the reality of the concept of turbulent viscosity for solar convection was
proposed by  \cite{schwarzschild59} to explain the first observations of solar 
granulation, arguing that the Reynolds number of these motions must be ${\cal O}(1)$ 
\citep[see also][]{stothers00,canuto00}. 

Of course, dissipation is not the only contribution of turbulence, which poses
a problem for DNS.  Given the restricted numerical resolution,  suitable values for 
the dissipation coefficients,  i.e., dynamic viscosity ($\nu$),  heat conduction 
($\kappa$) or magnetic diffusivity ($\eta$),  in hydrodynamic and magnetohydrodynamic 
simulations of convection, are typically orders of magnitude larger than the theoretical 
estimations of collisional transport coefficient in gas and plasma. Even though in DNS the 
explicit values of these parameters approach those estimated for turbulent dissipation coefficients, 
the results do not seem to achieve this asymptotic regime. For instance, 
\cite{2016ApJ...818...32F} performed DNS of solar convection in spherical 
shells progressively increasing the Rayleigh ($\Ra$) number. While they found that the energy
becomes independent of the heat conduction after a certain value of the Rayleigh number $\Ra$, 
neither the spectral distribution of energy nor the radial profile of the vertical velocity
indicated numerical convergence. In other words, it is unknown to what extent the numerical 
models of this phenomenon represent reality. 

An alternative to DNS is the large eddy simulations approach (LES). In LES, all the scales 
down to the numerical cut-off scale are simulated, while the action of unresolved scales is 
parametrized using the simulated values and turbulence scaling laws. It is a well-founded 
assumption when the unresolved turbulent flow is scale-invariant. However, this condition is 
not entirely fulfilled in convective systems with strong density 
stratification, rotation and/or other factors making the motions anisotropic 
\citep{2002IJNMF..39..855E}.
By using implicit large-eddy simulations (ILES), the contribution of the unresolved 
scales can be modeled by specially designed finite-difference truncation terms of 
the numerical advection.
In the  ILES methodology these terms act
as an effective dissipation, e.g., effective viscosity for the transport of
momentum, and as an energy flux between scales \citep{margolin+02}.

The numerical convergence of ILES simulations is not easy to define.
Since the terms in ILES would be truncated at a different scale, 
changing the spatial resolution of the simulation would necessarily affect the results.
Thus, the numerical convergence in such a situation is scale-dependent, as for
increasing numerical 
resolution the large scale properties of the flow remain unaltered while more and 
more small scale structures develop. 
Therefore, in this respect, the ILES approach is different 
from the LES, and requires a detailed investigation of the numerical convergence for particular 
classes of models.

This study explores the numerical convergence in ILES
2D simulations of turbulent stratified convection, which mimics the solar convection zone. 
We do not aim to compare the results to any particular analytical
solution or observations of a physical system, but explore whether
the integral characteristics of a numerical model reach a converged
state when the resolution increases. The initial and boundary conditions of our model 
resembles the solar convection zone, turbulent characteristics of which are still uncertain
and currently debated \citep{hanasoge+16,greer+15}. 
In this work
numerical convergence will be evaluated in terms of the vertical profiles of 
temporally and horizontally averaged quantities, like the turbulent velocities
and the convective heat flux.  The distribution of energy along the different 
spatial scales resolved for each grid is assessed through the turbulent
spectra of the kinetic energy and the variance of potential temperature.  
An approximated analysis of the effects of the ILES 
method on the dynamics of the system is performed from the estimation of an effective 
viscosity as a function of the scale. We compare this quantity with the turbulent viscosity
which is a rough measure of the enhancement of transport of momentum and other physical
quantities by turbulence. It is computed here from the length and time 
scales of the most energetic motions. Finally, the unexpected development
of oscillatory mean horizontal motions in the stable layer will also be presented.

A similar convergence study was performed by \cite{porter+94}, 
who used the ILES scheme based on a piecewise parabolic method \citep[PPM][]{colella+84}, 
devoid of explicit viscosity in the momentum equation, and
reached resolutions up to $1024\times256$ grid 
points (corresponding to $\Re\sim2\times10^4$). They found that at lower resolutions the vertical size of the
convection cells occupy the entire convection zone. For the 
highest resolution models, the small eddies break down the large cells 
resulting  in a flow dominated by structures of all sizes but without large-scale 
convection cells.  In the 3D simulations, they modified the upper 
thermal boundary condition allowing for the temperature to be constant in space 
but change with time \citep{porter+00}. They observed convergence in the results for
their high-resolution cases. \cite{sullivan+11} presented a 
convergence analysis for 3D LES simulations of the Earth convective boundary layer. They found
that numerical convergence is achieved whenever there is enough scale separation between
the most energetic eddies and those with scales close to the cutoff of the LES 
scheme. Besides the global convection simulations of \cite{2016ApJ...818...32F}
described above, to our knowledge DNS of stratified 
turbulent convection have not explored the role of resolution.

Three-dimensional simulations provide a better representation of 
the convection zone dynamics. For instance, in the presence of rotation 
and magnetic fields, the collective 3D effects of 
turbulence include the generation of large scale flows and dynamos. 
Most of these effects either do not exist or take a different form in 
2D convection. Nevertheless,  the influence of the finest scales on the 
largest scales is worth exploring, as it sheds light on turbulent convection 
as well as on the capabilities of ILES. In this work, we focus on 2D convection 
which allows for higher resolutions, while leaving 3D simulations for future studies.

This paper is organized as follows. In Section~\ref{sec:model}, we describe the equations, 
the numerical model, and other ingredients used to simulate turbulent convection.
In Section~\ref{sec:results}, we perform the convergence analysis by varying
the numerical resolution. Finally, in Section~\ref{sec:conclusions}, we present our concluding 
remarks.

\section{Numerical model}
\label{sec:model}

We used the EULAG-MHD code  \citep{SC13}--a specialized variant of the 
original EULAG code \citep{prusa2008eulag}--to perform two-dimensional convection 
simulations in a rectangular domain. EULAG-MHD is based on the multidimensional positive-definite 
advection transport algorithm, MPDATA \citep{smolarkiewicz2006multidimensional}.
It is a non-oscillatory forward-in-time advection solver with second-order accuracy in
space and time.  
The code allows simulations to be run as ILES
without any explicit dissipation (note that it also may be used for DNS with 
explicit dissipation). 
In the current model setup the vertical coordinate, $z$, 
spans from 0 to $L_z=254$ Mm (covering most of the depth of the solar convection zone and upper layers 
of the radiative zone), while  
the horizontal coordinate, $x$, spans from $0$ to $L_x = 2.5L_z$. The number of grid 
points is the same in the vertical and horizontal directions in each simulation.  
We solved the following set of Navier-Stokes 
equations governing mass, momentum and energy conservation,
\begin{equation}
    \nabla\cdot\rho_r\textbf{u}=0,
    \label{eq_mass_cons_nc}
\end{equation}
\begin{equation}
    \frac{d\textbf{u}}{dt} =
    - {\nabla \pi^{\prime}}
    - {\textbf{g}\frac{\Theta^{\prime}}{\Theta_r}} ,
    \label{eq_momentum_nc}
\end{equation}
\begin{equation}
    \frac{d\Theta^{\prime}}{dt} = -\textbf{u}\cdot\nabla\Theta_a - \alpha \Theta^{\prime}
    \label{eq_energy_nc}
\end{equation}

\noindent where $d/dt = \partial/\partial{t} + \textbf{u}\cdot\nabla$, $\textbf{u}$ is 
the velocity field, $\rho_r$ is the reference state density, 
which in the anelastic approximation is a function of the vertical coordinate 
only \citep{LH82}; $\pi^{\prime}$  
is the density normalized pressure perturbation, $p^{\prime}/\rho_{r}$; 
$\textbf{g}= -g \hat{z}$ is the gravity acceleration adjusted to fit the
solar gravity profile, and $\Theta$ is 
the potential temperature defined as $\Theta=T\left(P_b/P\right)^{R/c_p}$, where $T$ is the 
temperature, $P$ is the pressure, 
$P_b$ is the pressure at the bottom of the domain, $R$ is the universal gas 
constant, and $c_{p}$ is the specific heat 
at constant pressure. The potential temperature is equivalent to the specific entropy through the 
relation $ds = c_{p}d(\ln \Theta)$. The subscripts $r$ and  $a$ refer to the 
reference and ambient states, and 
the superscript $\prime$ means perturbations of a quantity around the ambient profile. 
Perturbations of potential temperature are related to perturbations of temperature by 
the anelastic approximation to the equation of state,  
$T^{\prime} = \Theta^{\prime} T_{a} / \Theta_{a}$. 
The energy equation contains a term forcing the adiabatic perturbations about the 
ambient state and a thermal relaxation term that dampens these perturbations in a 
inverse time scale $\alpha=1/\tau$.
In this setup, while the forcing term tries to mix the fluid in 
the convection zone, the thermal relaxation keeps the convective unstable state 
in this layer
\citep[see][for a comprehensive analysis of these effects]{cossette2017magnetically}.

\begin{figure}%
    \begin{center}
    \hspace*{-0.3cm}
    \includegraphics[height=0.62\columnwidth]{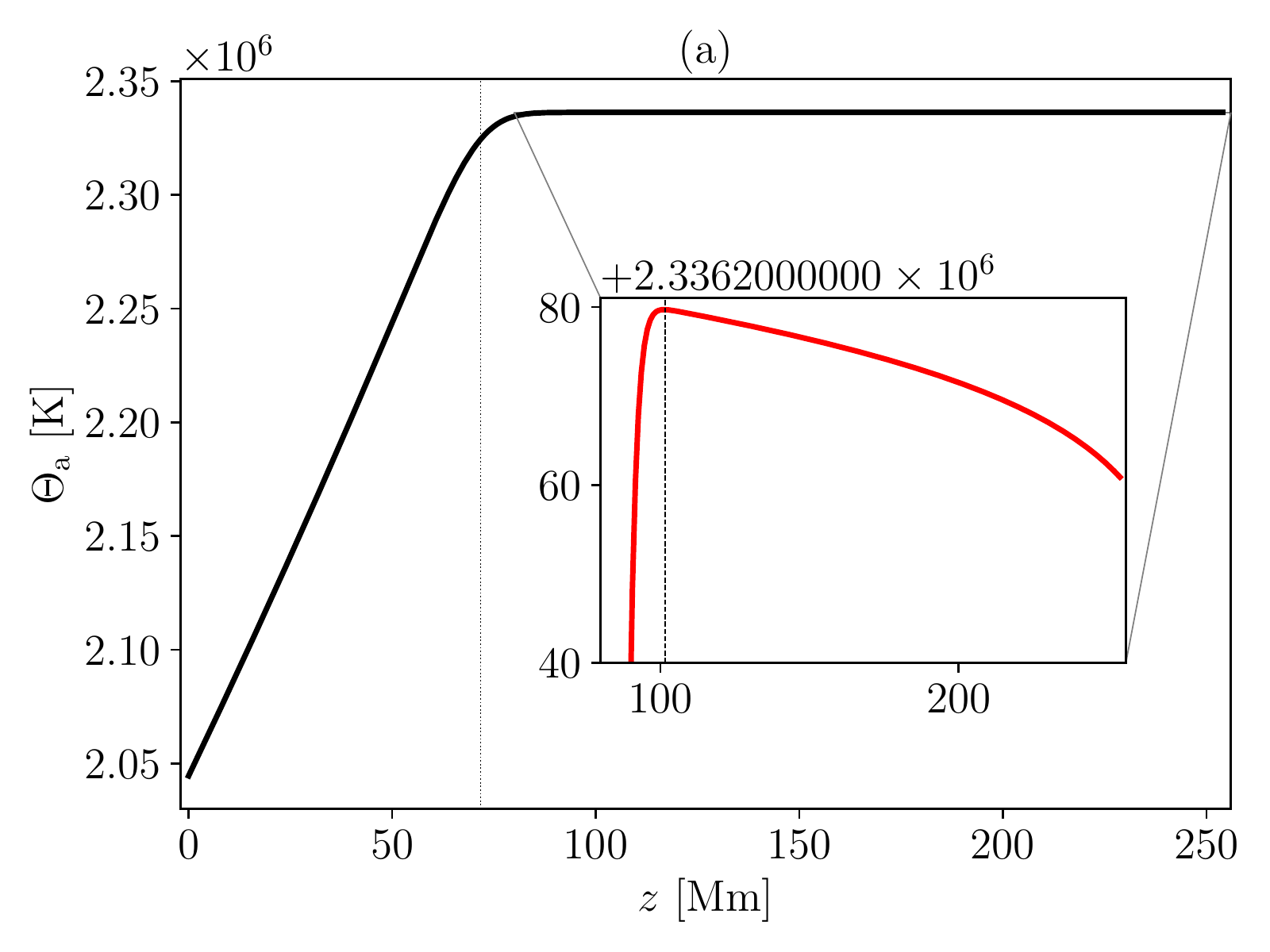} \\
    \hspace{0.3cm}
    \includegraphics[height=0.7\columnwidth]{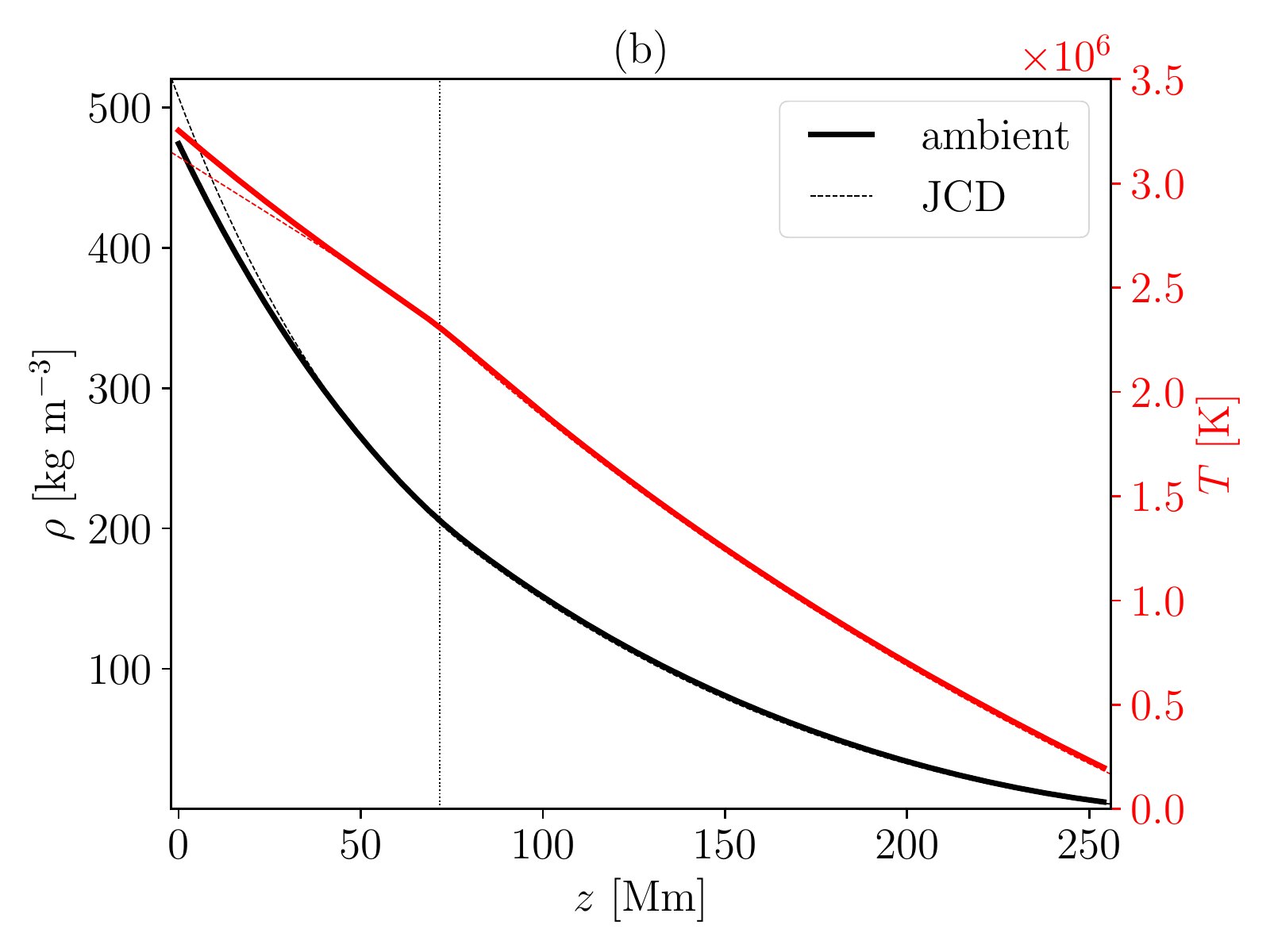} 
    \caption{(a) Vertical profile of the ambient potential temperature, $\Theta_a$.
    The insert shows a close-up for $0.1L_z < z < L_z$. (b) 
    Vertical profiles of the ambient density (black line) and temperature 
	(red line). The dotted lines correspond to the solar structure 
	model of \cite{CD+96}.}%
    \label{fig_ambient_state}%
    \end{center}
\end{figure}

The ambient state defining the thermodynamic variables, $\rho_a$, $\Theta_a$, and $T_a$ in 
Equations~\eqref{eq_mass_cons_nc}-\eqref{eq_energy_nc} is a particular solution of the 
hydrodynamics equations. In this work we construct the ambient state considering hydrostatic 
equilibrium as follows,  

\begin{equation}
    \frac{\partial T_a}{\partial z} = 
    -\frac{g}{R(m+1)},
    \label{est_amb1}
\end{equation}
\begin{equation}
    \frac{\partial\rho_a}{\partial z} = 
    -\frac{\rho}{T_a}
    \left(\frac{g}{R} + \frac{\partial T_a}{\partial z} \right),
    \label{est_amb2}
\end{equation}

\noindent where $m=m(z)$ is the polytropic index.  Solutions of 
Equations~\eqref{est_amb1} and \eqref{est_amb2} with  $m\geq 1.5$ correspond to stable 
stratification, while solutions for $m<1.5$ correspond to convectively unstable states.

We use an ambient state with a stable layer at the bottom of the domain by 
setting $m_{s}=2.5$ for $z\le0.28L_z$, and a marginally unstable convection zone 
with $m_{u}=1.499991$ for $z > 0.28L_z$.  This is achieved by considering a radial 
profile of the polytropic index, 

\begin{equation}
m(z) = m_{s} - \frac{1}{2}(m_{s} - m_{u}) \left[1 + \text{erf}\left(\frac{z-z_{1}}{w}\right) \right] \;,
\label{eq_m}
\end{equation}

\noindent where the transition between zones of different $m$ is made through the 
error functions with $z_1 = 0.28 L_z$ and $w=0.041 L_z$. Equations~\eqref{est_amb1} 
and \eqref{est_amb2} are integrated numerically with $\rho_{z_1}=208$ kg/m$^3$ and
$T_{z_1} = 2.322 \times 10^{6}$ K at the interface between the stable and the unstable layers,
$R = 13732$ J K$^{-1}$ kg$^{-1}$ is the gas constant for a monoatomic hydrogen gas,  and $c_p=2.5 R$.
The pressure is computed via the ideal 
gas equation of state, $P_a=R \rho_a T_a$. The resulting vertical profile of $\Theta_a$ is 
shown in Figure~\ref{fig_ambient_state}(a). In the convective zone, the slope of $\Theta_a$ is slightly 
negative with respect to the $z$ coordinate as it can be seen in the figure insert. 
The negative slope of $\Theta_{a}$ ensures that this zone is unstable to convection, with 
the difference of $\Theta_{a}$ between the bottom and top of the convectively unstable 
layer being $18$ K. The reference potential temperature $\Theta_r = T_{z_1}$.
Finally, for all the simulations, we have considered 
$\alpha = 1/\tau = 1.29\times10^{-8}$ s$^{-1}$.

The stable zone at the bottom of the domain ensures a more realistic transition
between the two layers, allowing a certain amount of overshooting.
This closely resembles situations in nature where convection happens in contact 
with a stable but not totally rigid 
layer. As we show below, the dynamics in this stable layer is governed by internal 
gravity waves (GW), which induce horizontal motions
and react back on the convection properties. 
Figure~\ref{fig_ambient_state}(b) shows the vertical profiles of the density, $\rho_a$, and
the temperature, $T_a$. The domain encompasses $4.5$ density scale heights.

The boundary conditions for this setup are defined as follows. 
In the horizontal direction, we consider periodic boundaries 
for all variables. In the vertical direction, stress-free and rigid boundary 
conditions for the velocity field are considered respectively at the bottom and 
top of the domain.
Null convective radial flux is considered as thermal boundaries at the bottom
and top as it has been used in previous works in the literature 
\citep[e.g.,][]{fan+03,2015ApJ...798...51H}.  The initial conditions are 
white noise correlated perturbations in both, $u_z$ and  $\Theta^{\prime}$, introduced
only in the unstable layer and with amplitudes $5\times10^{-4}$ m/s and $5\times10^{-4}$ K, 
respectively.  In the next section, we explore the
properties of physical quantities (in the real and spectral space) resulting from  
the simulations with the setup described above but for
different resolutions, namely $N=64$, $128$, $256$, $512$, $768$, $1024$, 
and $2048$ cells in each direction. All the simulations were run for at least 60 years.
This time is sufficient for the simulations to reach a statistically steady state, and
to provide enough data for the analysis.
The parameters and results of the simulations
are summarized in Table~\ref{table_results}.

\begin{deluxetable*}{ccccccc}
\tablenum{1}
\tablecaption{Simulation parameters and results \label{table_results}}
\tablewidth{0.9\columnwidth}
\tablehead{
\colhead{Simulation} & \colhead{$N$} & \colhead{$\langle \urms \rangle$  [m/s]} & 
\colhead{$\overline{\nu}_{\rm eff}$ [$\times 10^9$  m$^2$/s]} & \colhead{$\Re_{\rm eff}$} 
	& \colhead{$\ell$ [Mm]} & \colhead{$\nut$ [$\times 10^9$ m/s$^2$ ]} 
}
\startdata
FD64   &   64   & 34.0 & 9.8983 & 0.7174 & 42.7 & 0.484 \\
FD128  &   128  & 37.5 & 0.9820 & 7.9923 & 46.5 & 0.583 \\
FD256  &   256  & 40.4 & 0.0764 & 110.56 & 48.2 & 0.651 \\
FD512  &   512  & 41.9 & 0.0115 & 758.77 & 52.3 & 0.732 \\
FD1024 &   1024 & 43.5 & 0.0033 & 2698.7 & 51.8 & 0.754 \\
FD2048 &   2048 & 43.8 & 0.0013 & 6933.6 & 51.2 & 0.749 \\\hline
\enddata
\tablecomments{Results of simulations FD with different resolutions, $N$. 
In $\langle \urms\rangle$ the velocity is averaged in space and time,  
The effective viscosity, $\overline{\nu}_{\rm eff}$  is the average over 
the largest wavenumbers, $k$. 
The effective Reynolds number, $\Re_{\rm eff}=\langle \urms\rangle L \overline{\nu}_{\rm eff}$.
The convective correlation length, $\ell$, is calculated from averages of 
the kinetic energy spectra at different times according to Eq.~\ref{eq_corr_len}. 
Finally, $\langle u_{rms}\rangle$ and $\ell$ are
used to compute the turbulent viscosity, $\nut$, following Eq.~\ref{eq_turb_visc}. 
The spatial averages were calculated for the convectively unstable layer 
only ($0.3L_z <z<L_z$).}
\end{deluxetable*}

\section{Results}
\label{sec:results}

\subsection{Analysis in physical space} 
\label{sec:phys}

\begin{figure*}%
    \centering
    \includegraphics[width=0.9\columnwidth]{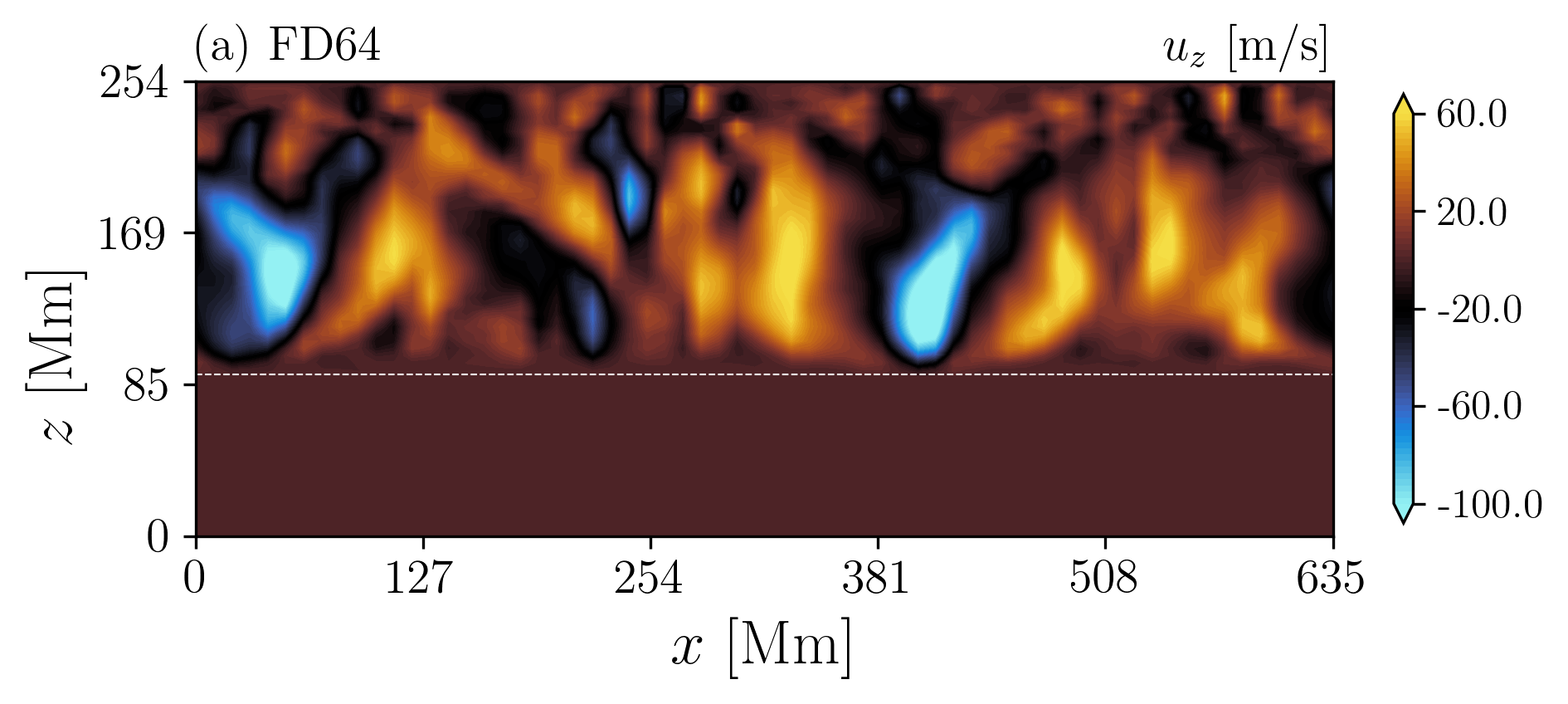} 
    \includegraphics[width=0.9\columnwidth]{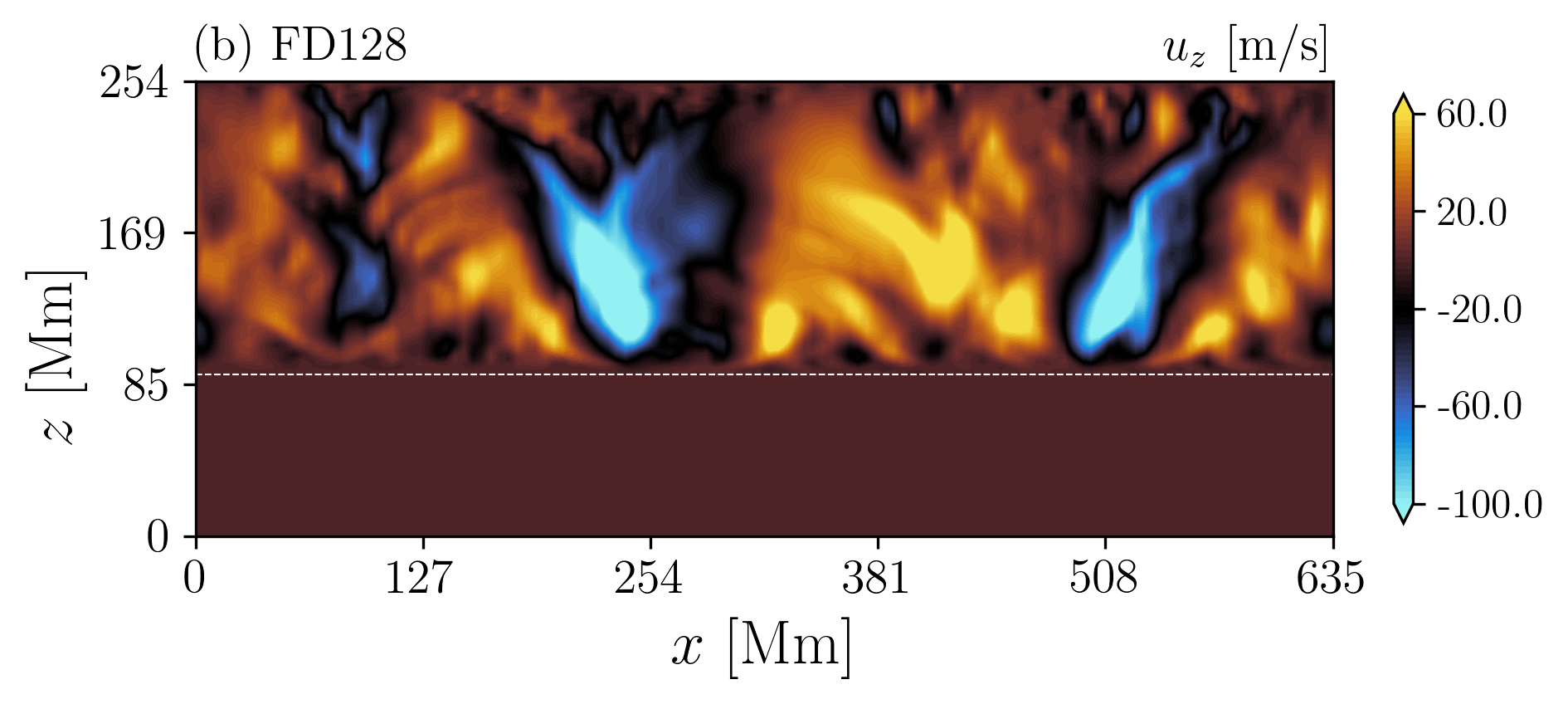} \\
    \includegraphics[width=0.9\columnwidth]{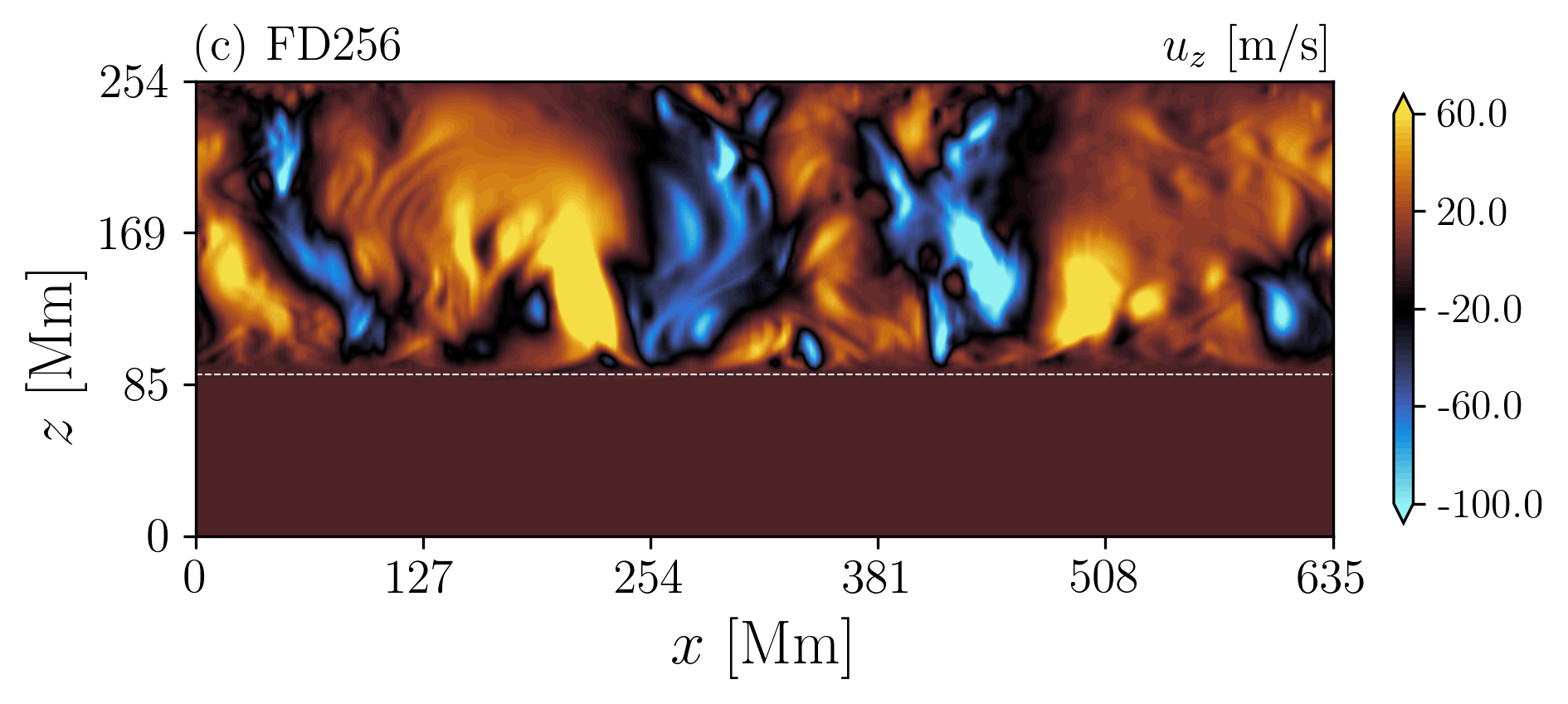} 
    \includegraphics[width=0.9\columnwidth]{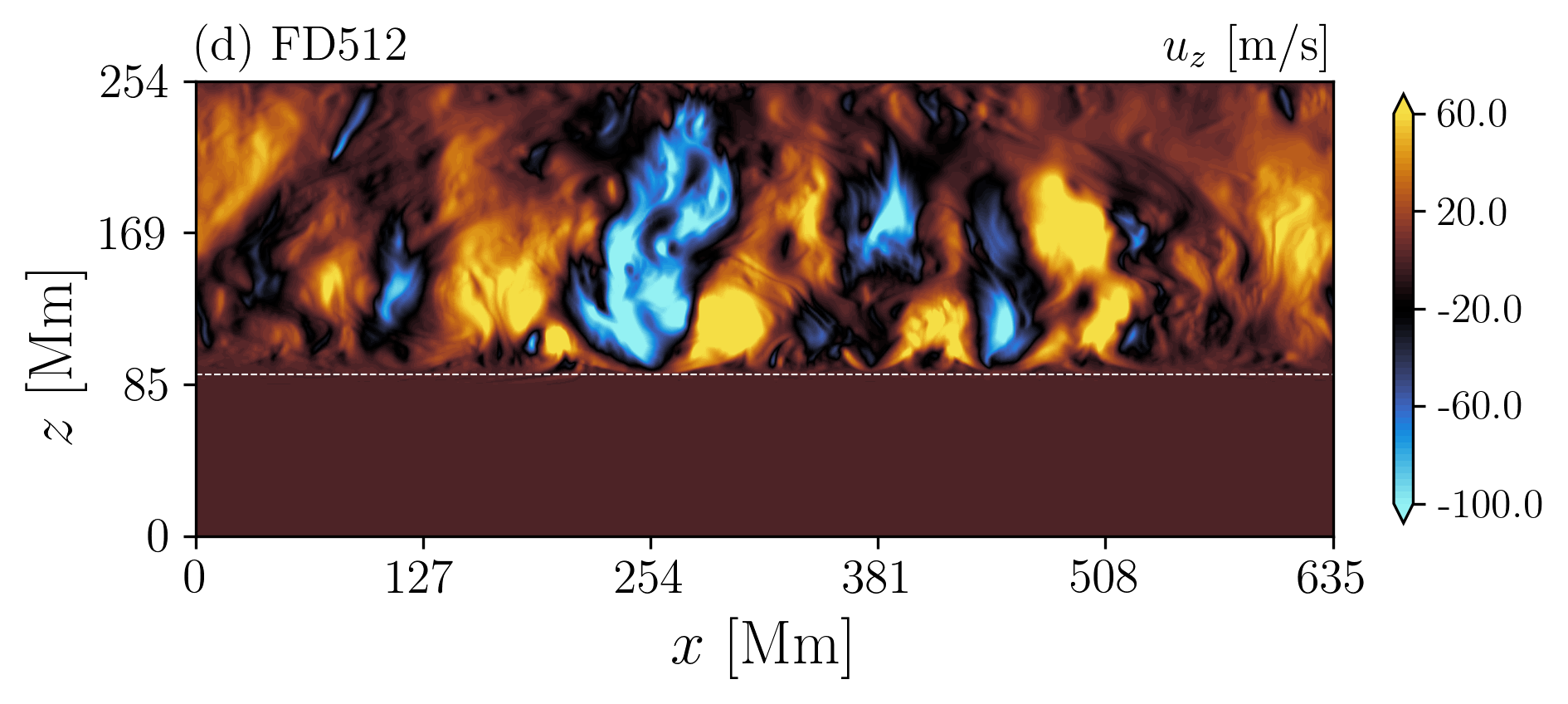} \\
    \includegraphics[width=0.9\columnwidth]{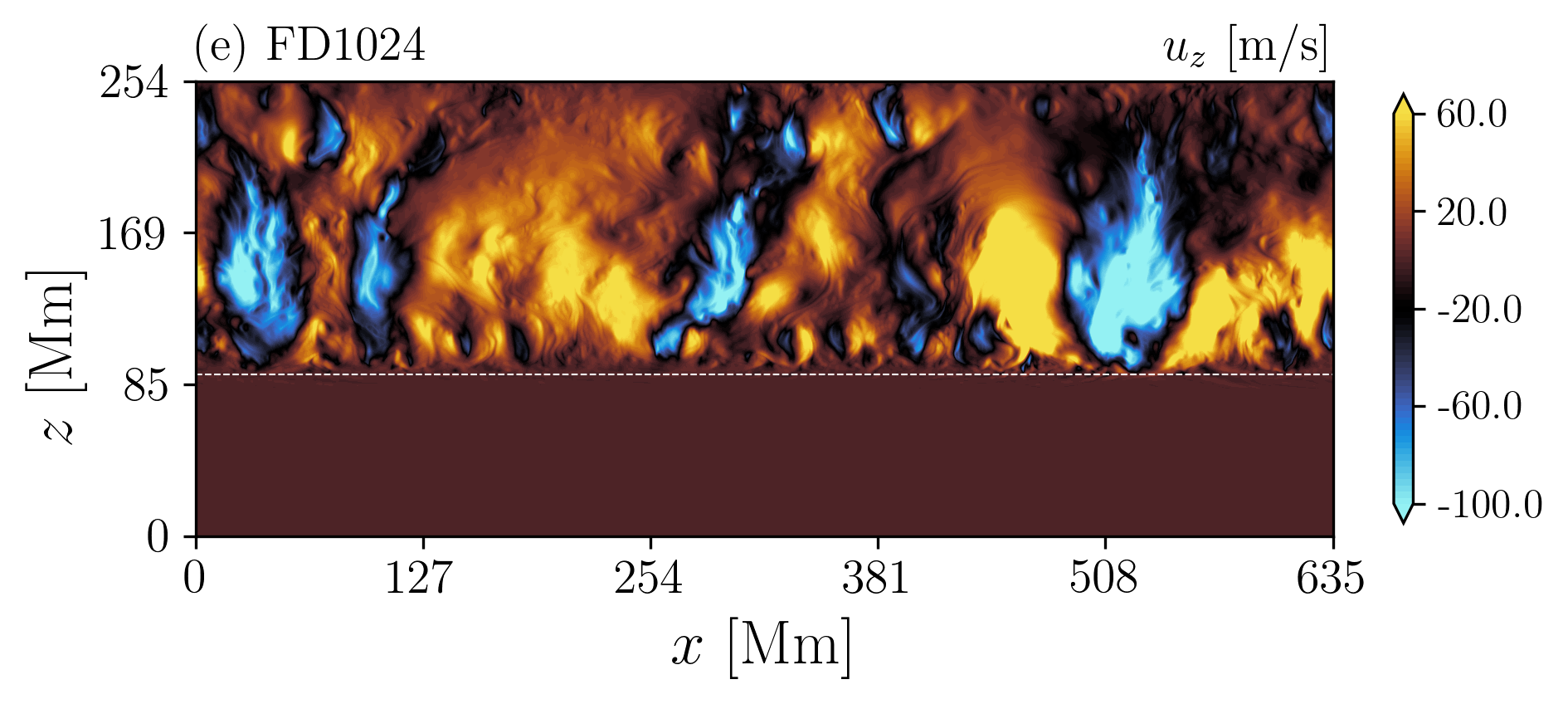}
    \includegraphics[width=0.9\columnwidth]{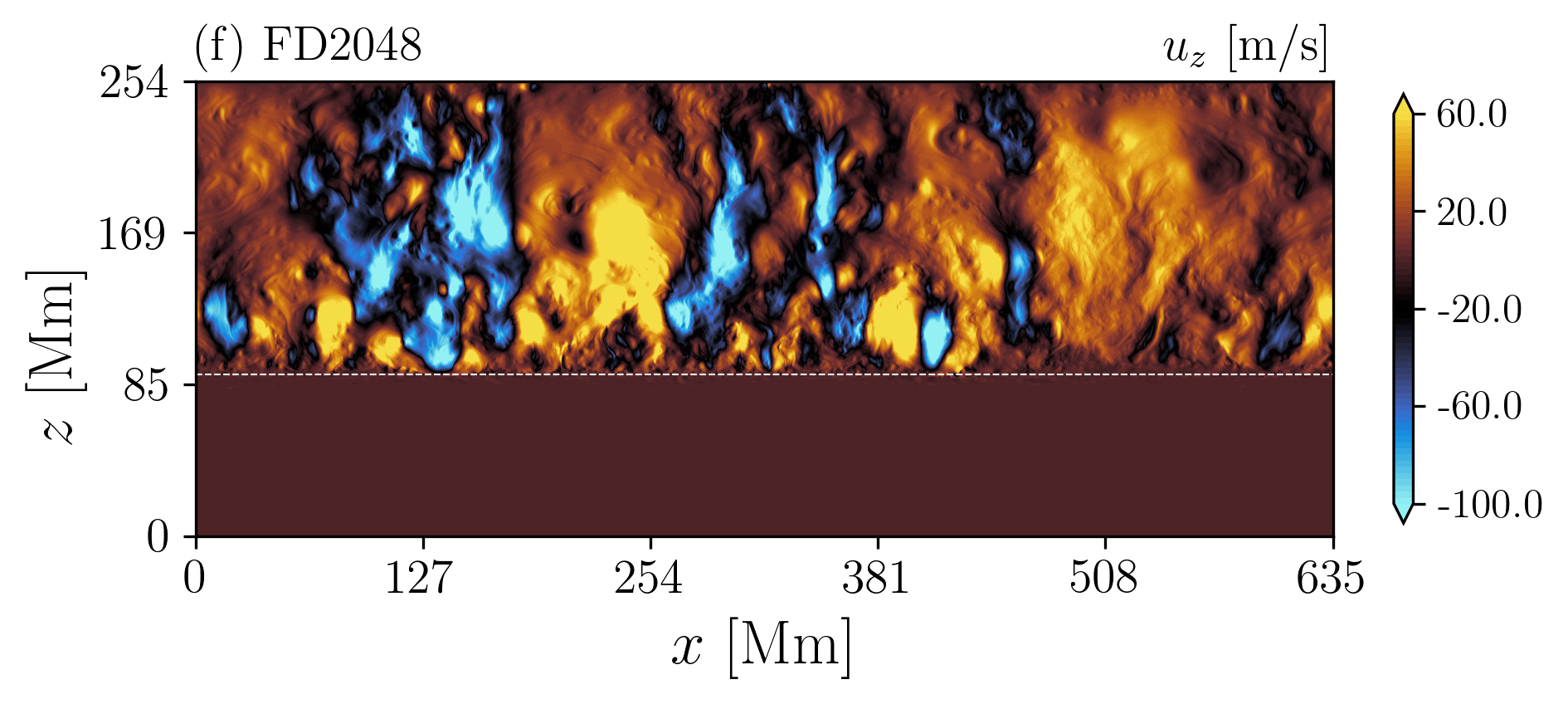}
    \caption{Snapshots of vertical velocity for simulations (a) FD64 to  (f) FD2048. 
	Yellow (blue) colors correspond to upflows (downflows). The label of each 
        simulation is shown on the top 
	of each image.}%
    \label{snapshots_w}%
\end{figure*}

Figure~\ref{snapshots_w} shows snapshots of the vertical velocity for simulation
(a) FD64,  (b) FD256, (c)  FD768 (c),  and (d) FD2048.  The yellow and blue contours
represent upflows and downflows, respectively.  The canonical picture of
convection in the environment with a unstably stratified density profile
describes broad upflows and narrow downflows. The figure shows that this
feature is well captured by the 2D convection model for low and high resolutions, forming
two or three large convective cells.  Unlike 
previous ILES results \citep{porter+94}, in the higher resolution simulations ($N\ge512$), these large 
cells  are not broken by the small-scale structures but coexist with them. The strong
downflows seem to be formed by the coalescence of thinner plumes observed at the 
upper part of the domain.

\begin{figure*}%
    \centering
    \includegraphics[width=1.8\columnwidth]{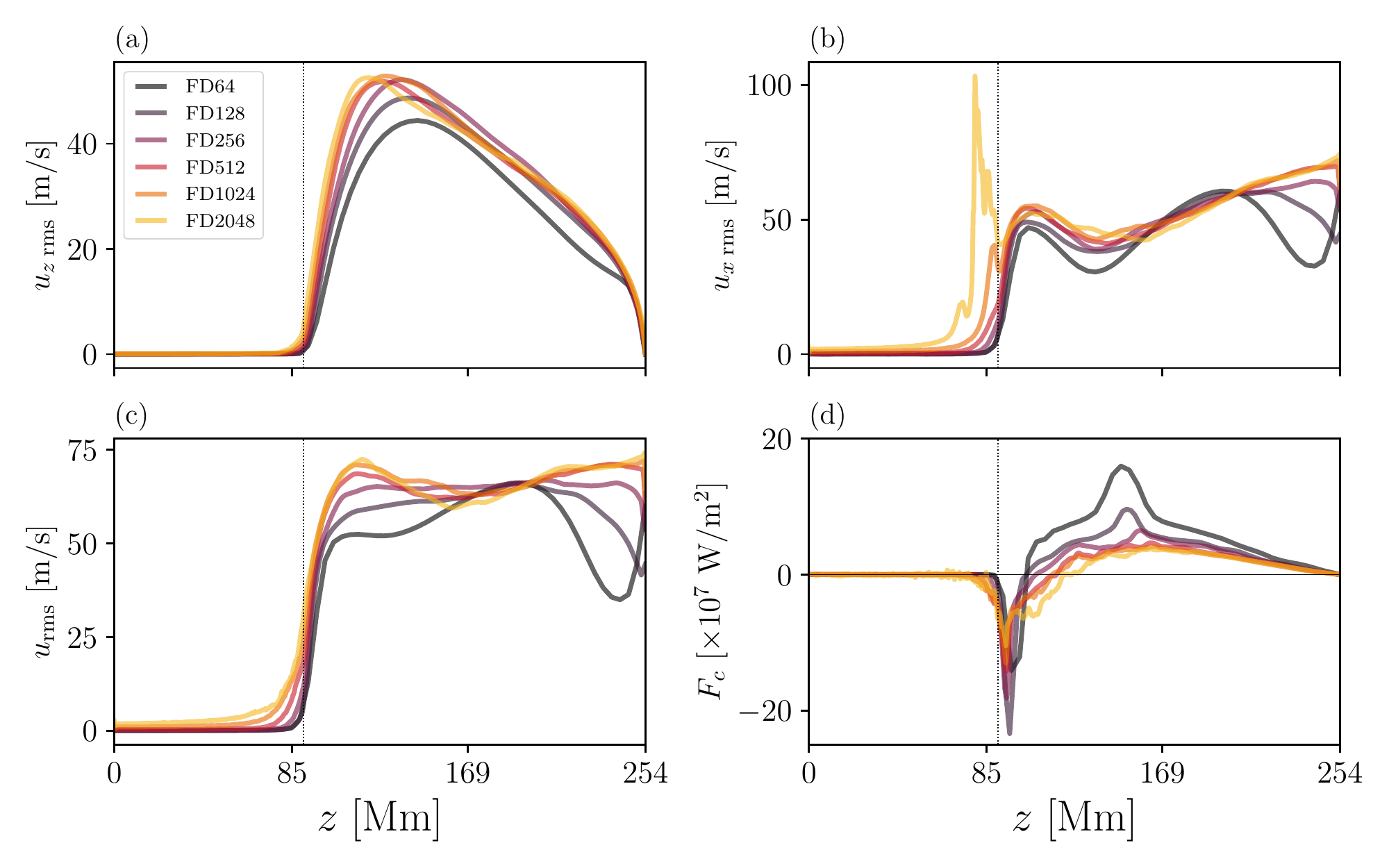} 
    \caption{Vertical profiles of the (a) vertical; (b) horizontal; (c) total rms 
    velocities; and (d) the convective heat flux, $F_c = \rho_a c_p \brac{u_z \Theta^{\prime}}$ 
    for simulations FD64-FD2048.  Different resolutions are represented by different 
    colors indicated in panel (a). }%
    \label{urms_nc}%
\end{figure*}

\begin{figure*}%
    \centering
    \includegraphics[width=0.9\columnwidth]{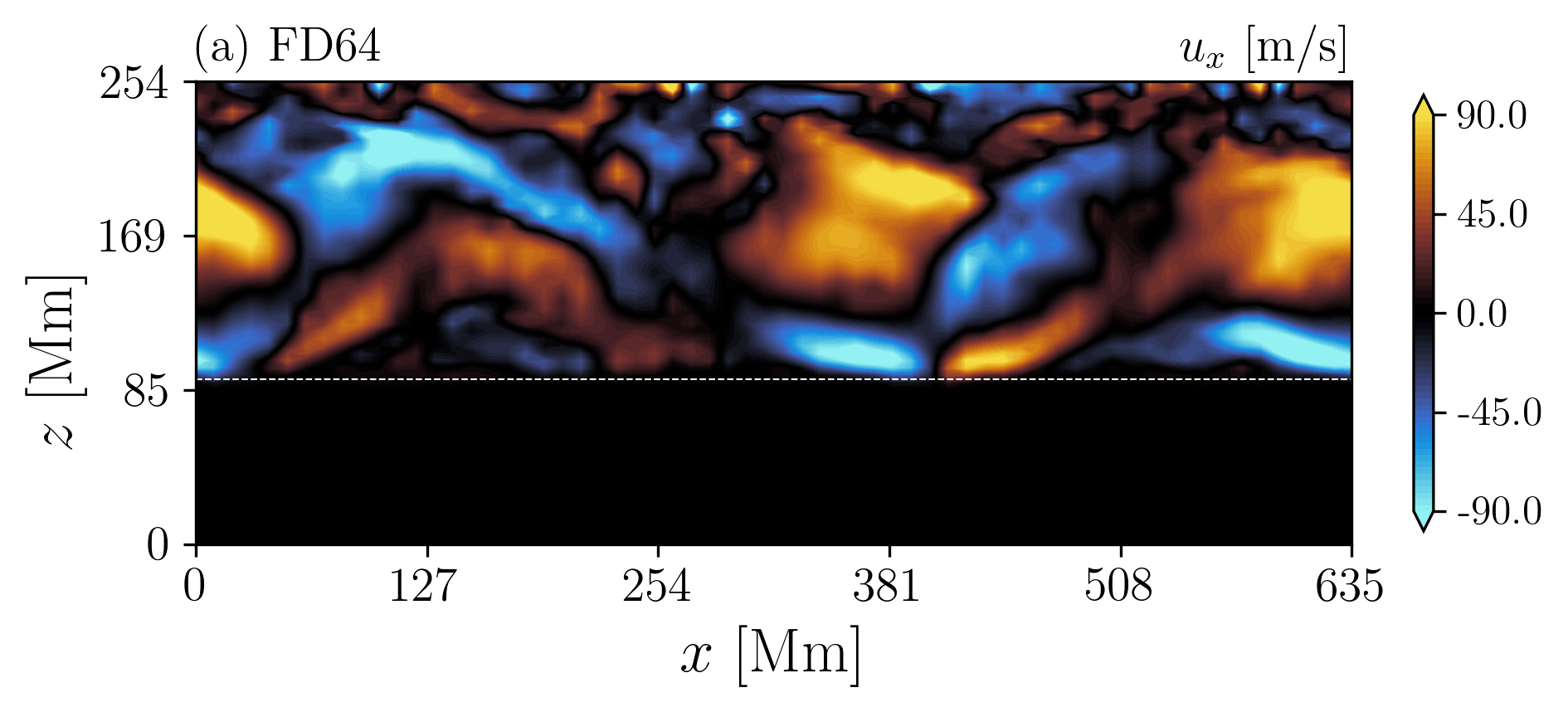} 
    \includegraphics[width=0.9\columnwidth]{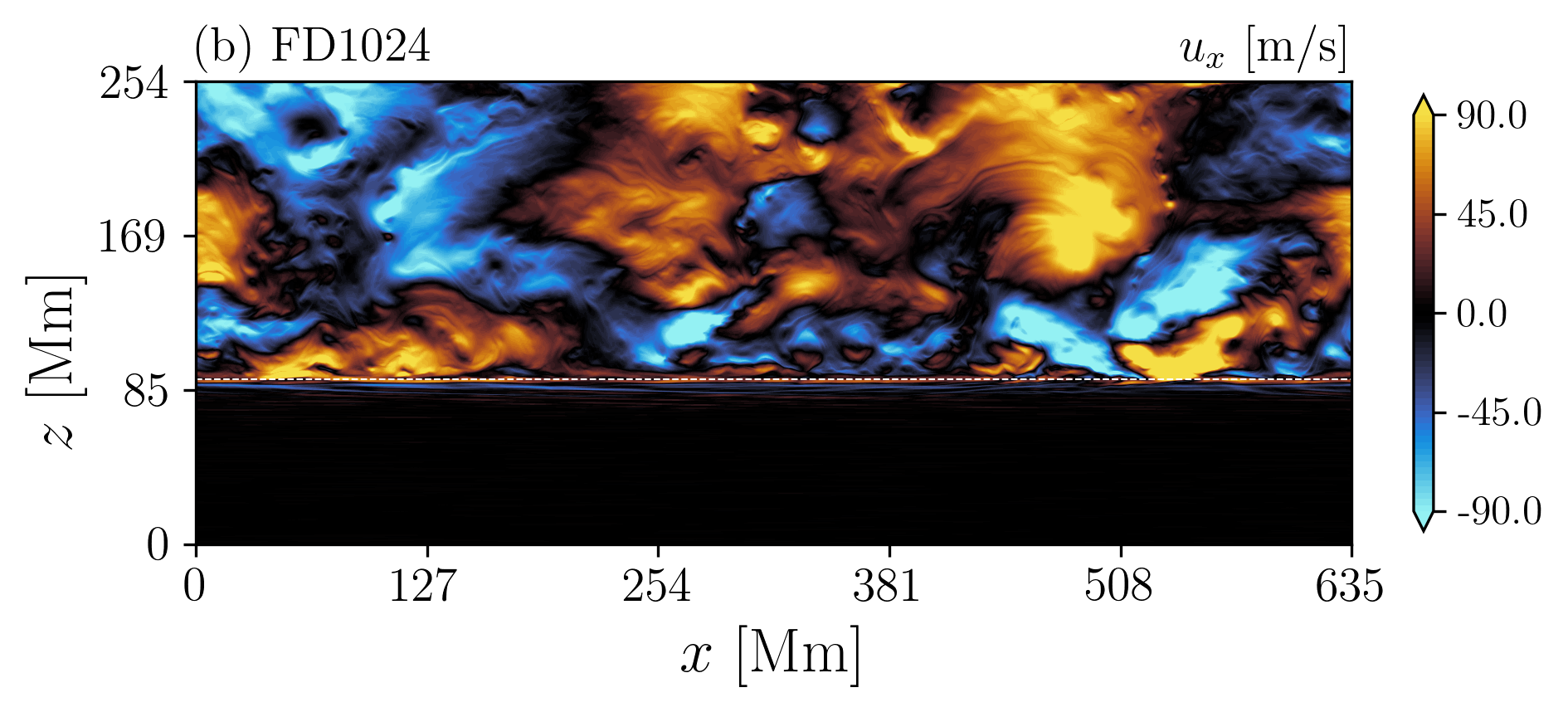} 
    \caption{Snapshots of horizontal velocities for simulations (a) FD64, and (b) FD1024.
	Yellow (blue) contours indicate horizontal flow towards the right (left).}%
    \label{snapshot_uxrms_nc}%
\end{figure*}

Figure \ref{urms_nc} shows $u_{z \; {\rm rms}}$ (a), $u_{x \; {\rm rms}}$ (b), 
$\urms=\sqrt{u_{x}^{2}+u_{z}^{2}}$ (c), and the convective flux, 
$F_c = c_p \rho_r \Theta^{\prime} w$ (d) as a function of the vertical coordinate $z$.
The rms profiles correspond to averages over the horizontal direction and time during
the last 15 years of the simulations.  
The profiles of panel (a) are consistent with similar simulations presented
in the literature for 2D simulations \citep{hotta2012numerical}, i.e., the vertical
velocity has a peak close to the bottom of the domain. In 3D simulations, compatible to
the ones presented here, this maximum is shifted towards the top 
\cite[e.g.,][]{chan1986turbulent,fan+03,hotta2012numerical}.  
Except for the amplitude, there is no significant change in this profile for different
resolutions, and simulations with $N \ge 512$ seem to have reached convergence in
the location and the value of the maximum value.  The profiles of horizontal 
velocity exhibit significant variations with resolution.  
For the low-resolution simulations ($N \leq 128$), $u_{x\> {\rm rms}}$ has two maxima, at the
upper and lower parts of the convective layer, and a minimum at $z\sim135$ Mm. The maxima 
correspond to fluid displacements in the opposite direction of the large eddies, whereas the 
minimum corresponds to the radius where the reversal takes place. 
In the lower part of the convection zone, this motion has a smaller vertical extension 
than in the upper part as a consequence of density stratification. 
When the resolution increases, the profile of $u_{x \; {\rm rms}}$  becomes flattened
with less prominent maxima and minima.  This can be explained by
the small scale structures developed in large resolution simulations, see Fig.~\ref{snapshot_uxrms_nc}(b) 
corresponding to simulation FD1024 as contrasted to the 
smooth large eddies observed in the low resolution simulations, see Fig.~\ref{snapshot_uxrms_nc}(a)
corresponding to simulation FD64. In the upper layers, the shorter density scale heights 
enforces smaller convective structures. For the simulation with $N=64$, there are 8 or less 
grid points per density scale height, which is insufficient to resolve small-scale convective 
motions, resulting in larger convective cells. Thus, the horizontal flows have another
reversal which on average forms a second minimum above $z\sim250$ Mm (see dark grey line in 
Fig.~\ref{urms_nc}(b) and also the upper part of Fig.~\ref{snapshot_uxrms_nc}(a)).
On the other hand, the simulations with $N > 512$ have at least 32 grid points per density scale 
height and are able to resolve these structures
with the appropriate energy 
(see, for instance, Figs.~\ref{snapshots_w}(d-f) and \ref{snapshot_uxrms_nc}(b)).
After the temporal and horizontal averaging, these cells are wiped out, and the minimum 
disappears.

Another interesting feature observed in the horizontal velocity is their depth of penetration
which increases with the numerical resolution. Additionally, 
there is a sharp peak evident in the cases FD1024 and FD2048, which is formed due to
generation of internal gravity waves in the stable layer. These waves start to become evident
in simulations with $N > 256$ whenever strong downflows induce perturbations in the 
stable zone (these perturbations are evident in Fig.~\ref{snapshot_uxrms_nc}(b), after 
some magnification of the figure, below the white dashed line). 
However, they are evanescent and dissipate on time scales depending on the numerical 
resolution. The waves appear to be resolved only
in the simulations with $256^2$ grid points or greater.
For the simulations FD1024 and FD2048, the effective viscosity is so small
(see \S\ref{sec:sgs}) that the perturbations induced by the gravity waves do not dissipate. 
Instead, they interfere, and upon a spontaneous symmetry breakdown interact non-linearly to 
form mean horizontal flows \citep{galmiche2000wave, wedi+06}, evidenced in Fig.~\ref{urms_nc}(b) by the peaks 
with magnitudes that increase with the resolution. 
For some resolutions these motions create oscillatory patterns; 
see \S\ref{sec:qbo} for a discussion.

\begin{figure*}%
    \centering    
    \includegraphics[width=0.94\columnwidth]{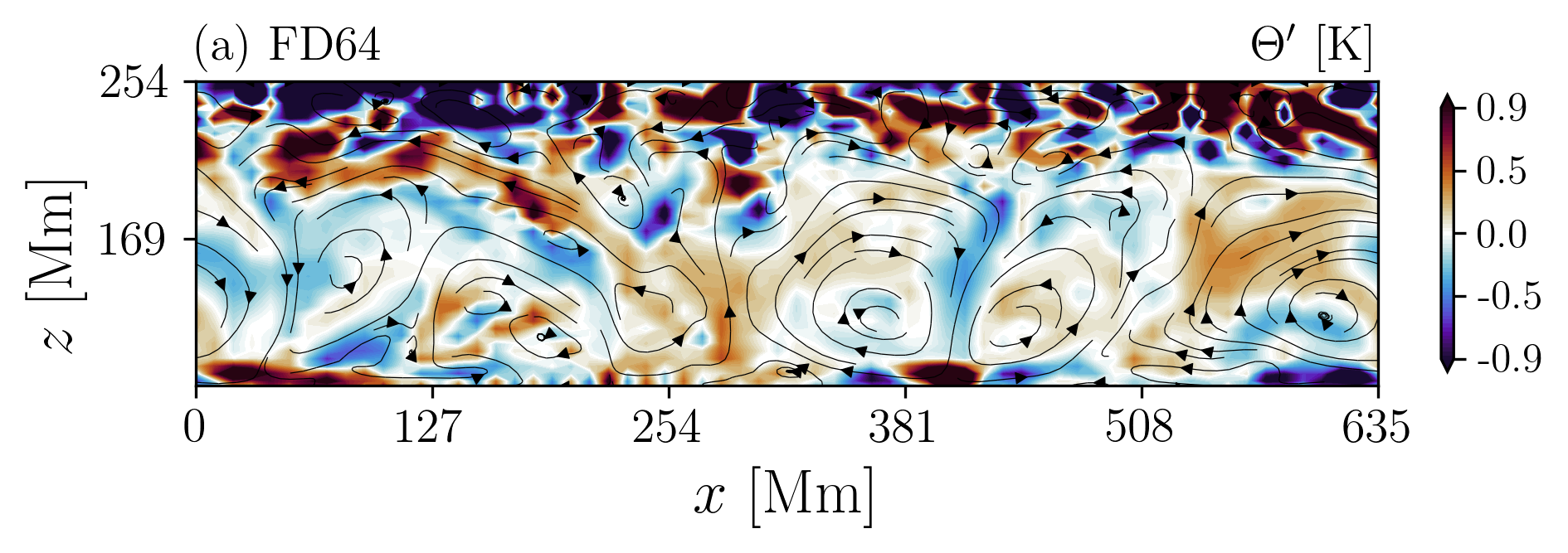} 
    \includegraphics[width=0.94\columnwidth]{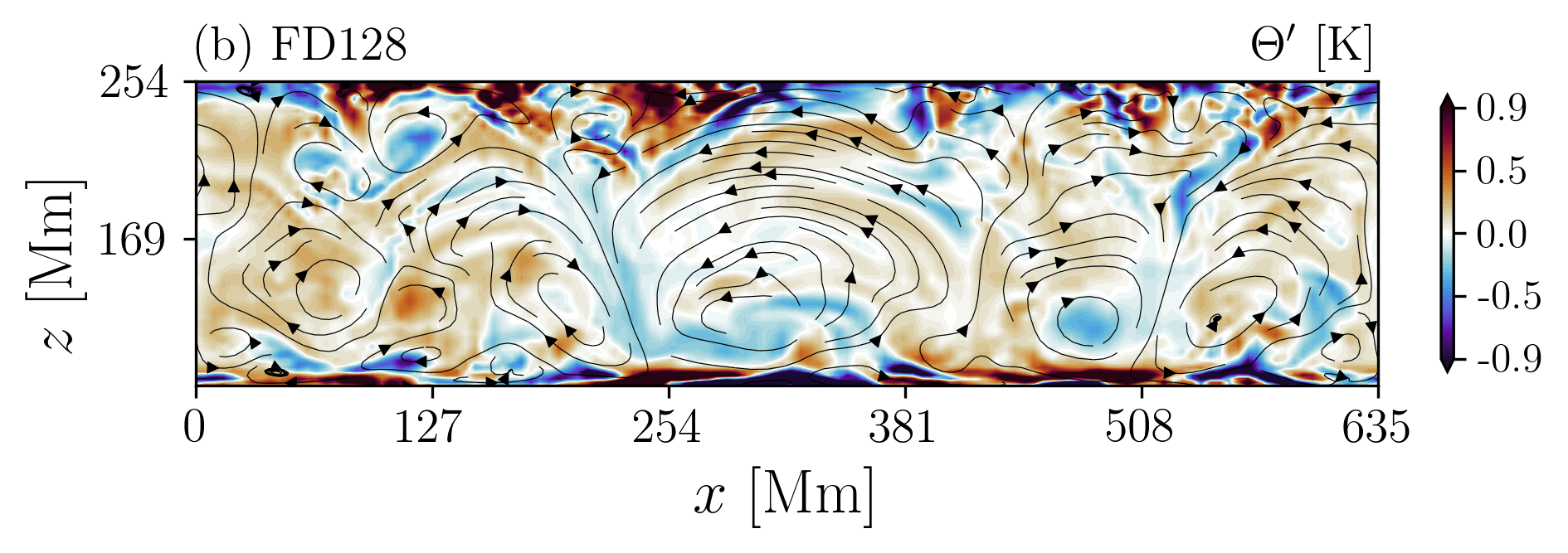}  \\
    \includegraphics[width=0.94\columnwidth]{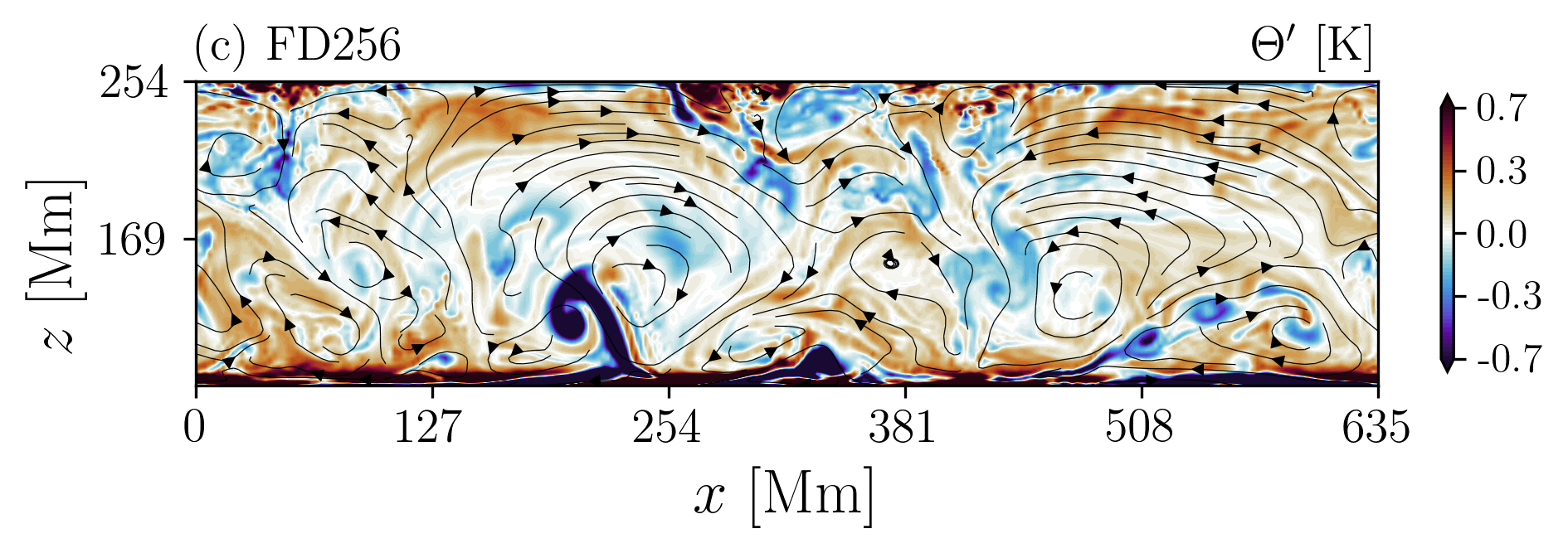} 
    \includegraphics[width=0.94\columnwidth]{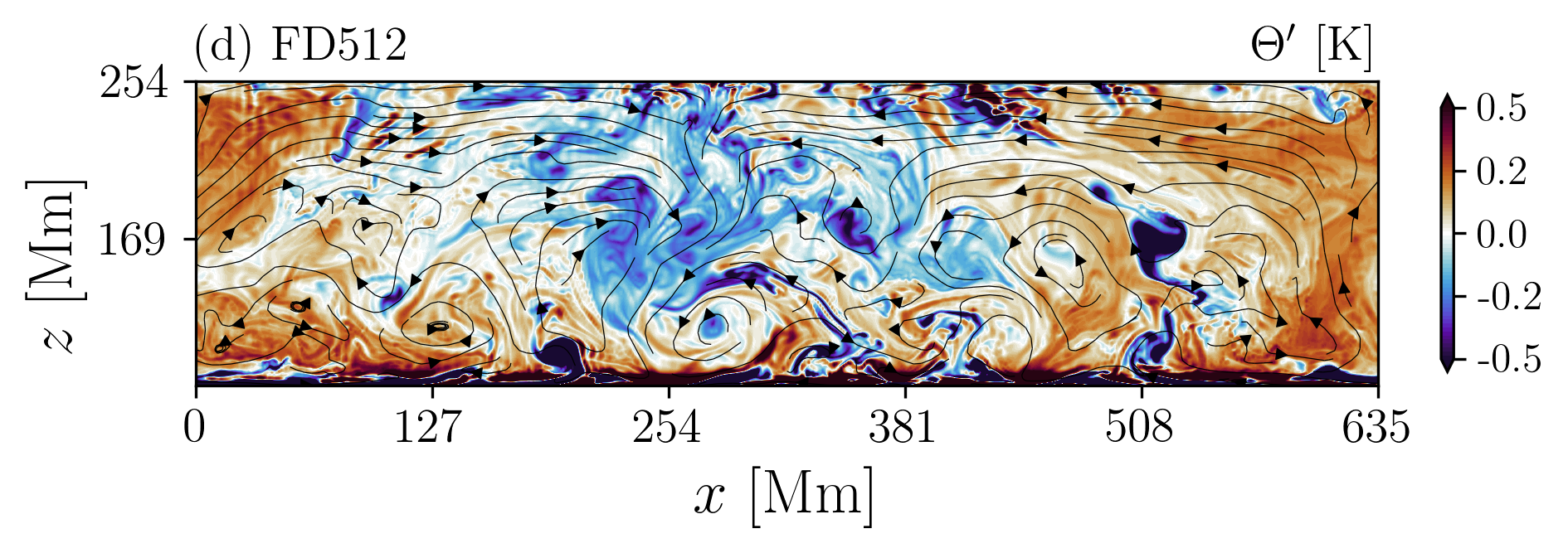} \\
    \includegraphics[width=0.94\columnwidth]{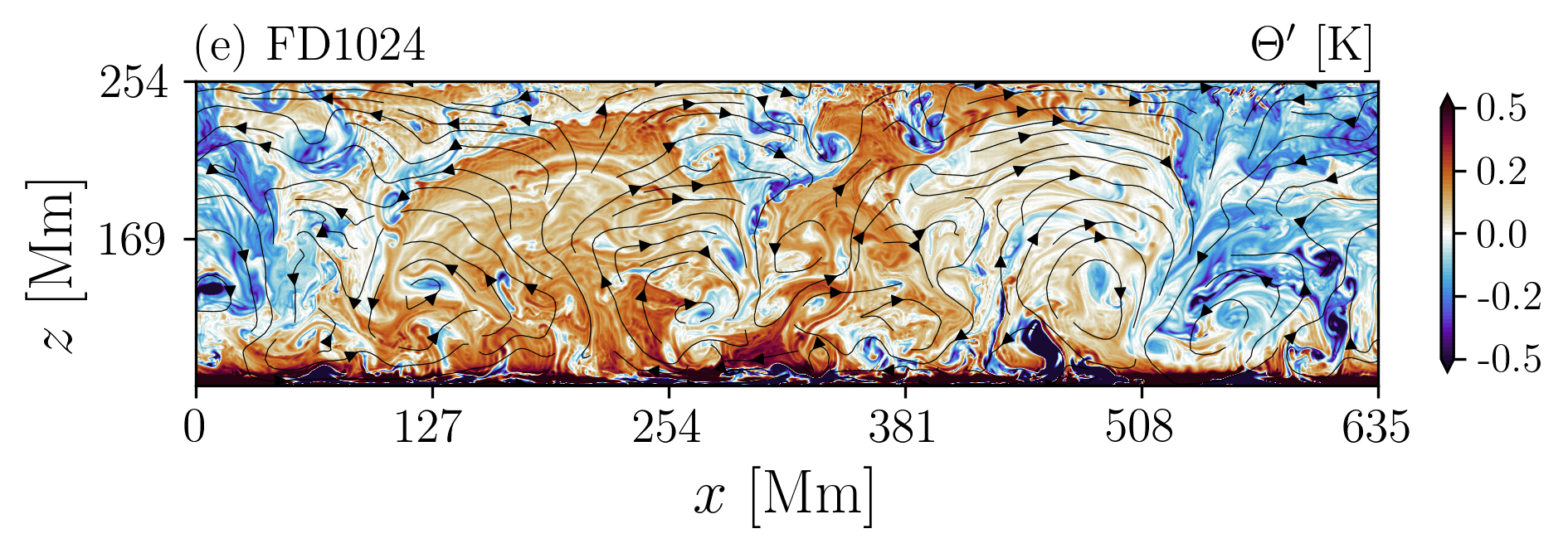} 
    \includegraphics[width=0.94\columnwidth]{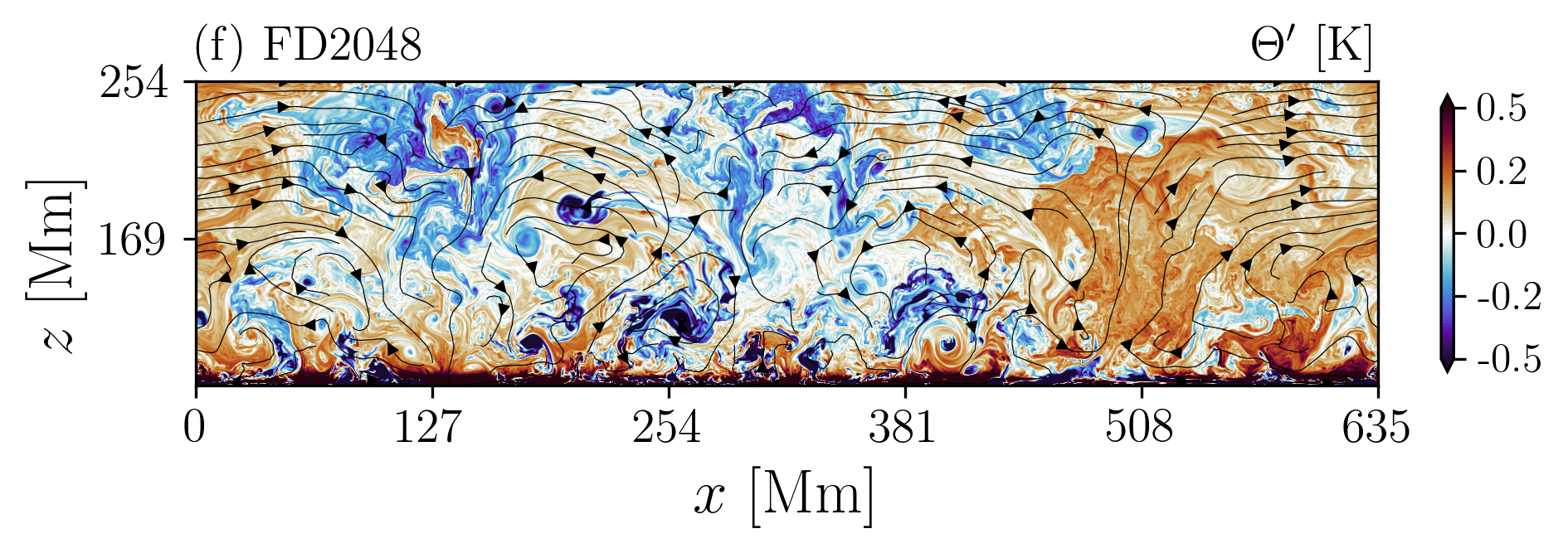} \\
    \caption{Snapshots of perturbations of potential temperature residual, 
    $\Theta^{\prime} - \brac{\Theta^{\prime}}_z$, for different resolutions. 
    Red contours indicate positive values (hot plumes), blue contours 
    correspond to negative values (cold plumes). The black streamlines show 
    the velocity field with the arrows pointing in the direction of the flow.}%
    \label{snapshots_the_nc}%
\end{figure*}

The profiles presented in Fig.~\ref{urms_nc}(c) are combinations of the corresponding motions 
depicted in panels (a) and (b). Their resemblance to the profiles in panel (b) reflects the fact 
that the horizontal motions are dominant in the convection zone. Finally, the convective heat flux presented 
in Fig.~\ref{urms_nc}(d) exhibits negative values at the bottom of the
convective layer identified with penetrative motions. The higher the resolution the larger
the extent of the overshooting region. On the other hand, the profiles with positive values 
in the convection zone show that the averaged heat flux carried by convection decreases 
with $N$ and seems to reach convergent values for $N>512$. 

Figure~\ref{snapshots_the_nc} shows snapshots of the residual of the
potential temperature perturbations,
$\Theta^{\prime} - \brac{\Theta^{\prime}}_z$, with the 
angular brackets meaning vertical average,  superimposed with streamlines of the 
velocity field for all simulations.
For a better contrast of the structures we have removed the stable layers from the figures.
Note that increasing the resolution allows for progressively more intricate structures in 
the form of filaments following the edges of convective eddies. These filaments form spikes 
and swirls due to the fluid movement until they dissolve.

\subsection{Analysis in the Fourier space}
\label{sec:spectral}

To explore further the turbulent characteristics of the convective motions 
we compute the kinetic power spectrum, $\tilde{E}^K(k,z)$, and the turbulent spectrum of the variance 
of $\Theta'$. The one dimensional, temporally averaged, kinetic power spectrum is defined as
\begin{equation}
    \tilde{E}^K(k, z) = \frac{1}{2} \tilde{u}(k, z)^{\ast}\tilde{u}(k, z),
    \label{eq_spectra}
\end{equation}

\noindent where $k$ is the wave number in the $x$ direction, the tilde denotes the Fourier transform of a 
quantity, and the asterisk denotes complex conjugate. The temporal average is performed considering 
the last 15 years of evolution with a sampling rate of 4 months. The power spectrum of the variance of 
potential temperature,  $\tilde{E}^{\Theta}(k, z)$,  
is defined analogously by replacing $\tilde{u}(k, z)$ by $\tilde{\Theta^{\prime}}(k, z)$ 
in Eq.~(\ref{eq_spectra}). Since we expect the properties of convection to depend 
on the height, we perform this computation for the top and the middle of the convection 
zone, $z=223$ and $z=139$ Mm, respectively. 

\begin{figure*}%
    \centering
    \includegraphics[width=0.87\columnwidth]{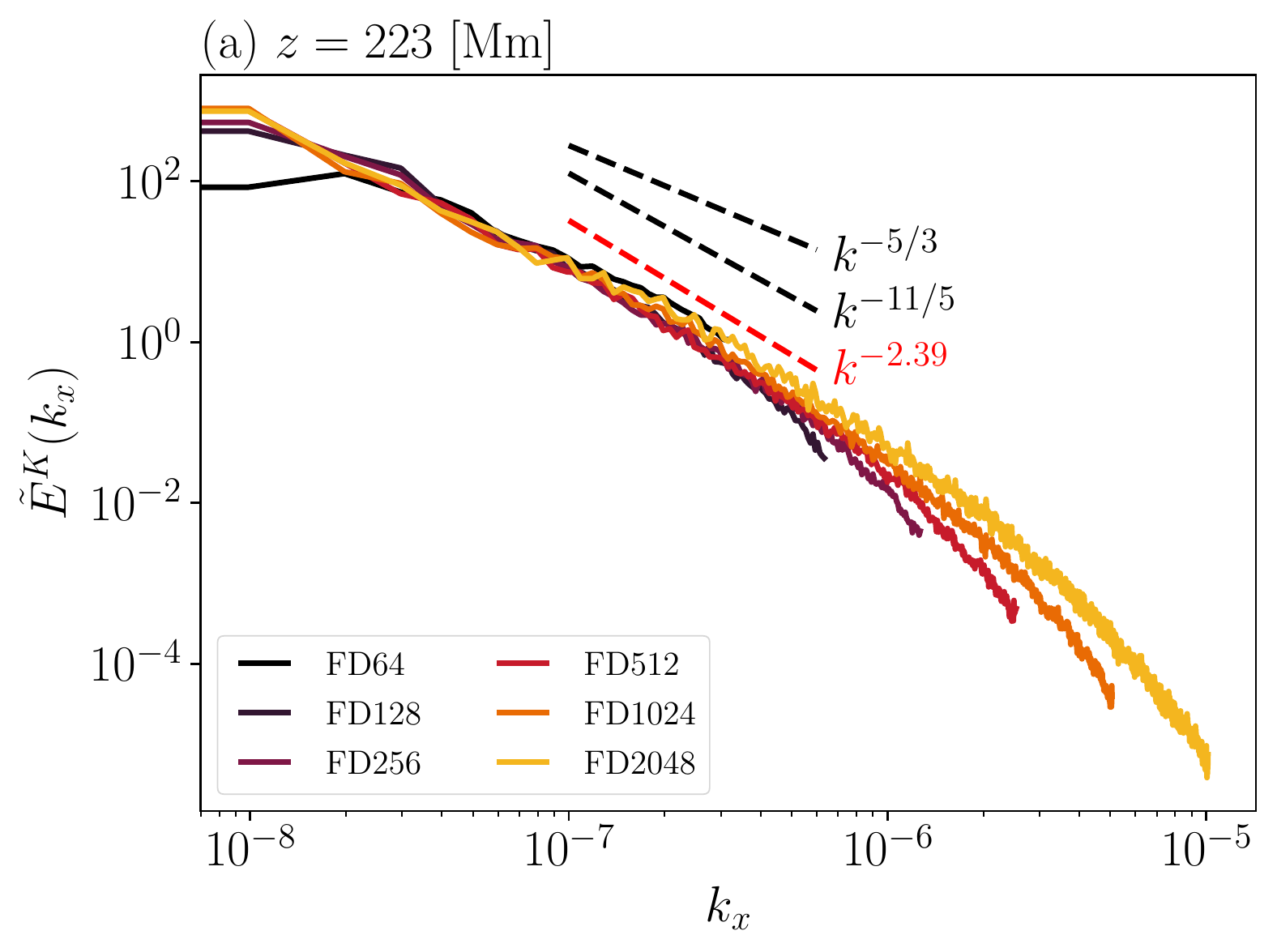} 
    \includegraphics[width=0.87\columnwidth]{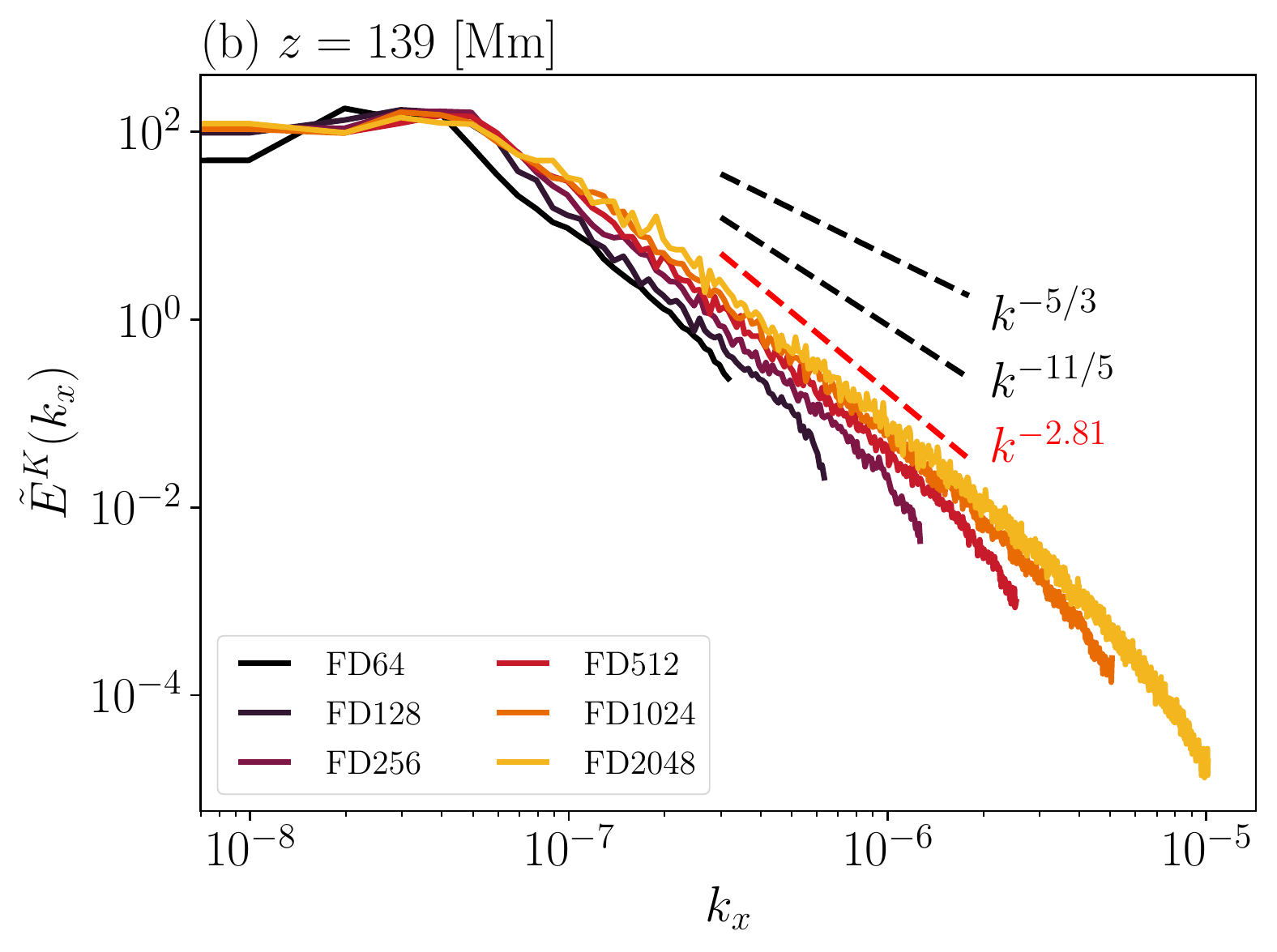} \\
    \includegraphics[width=0.87\columnwidth]{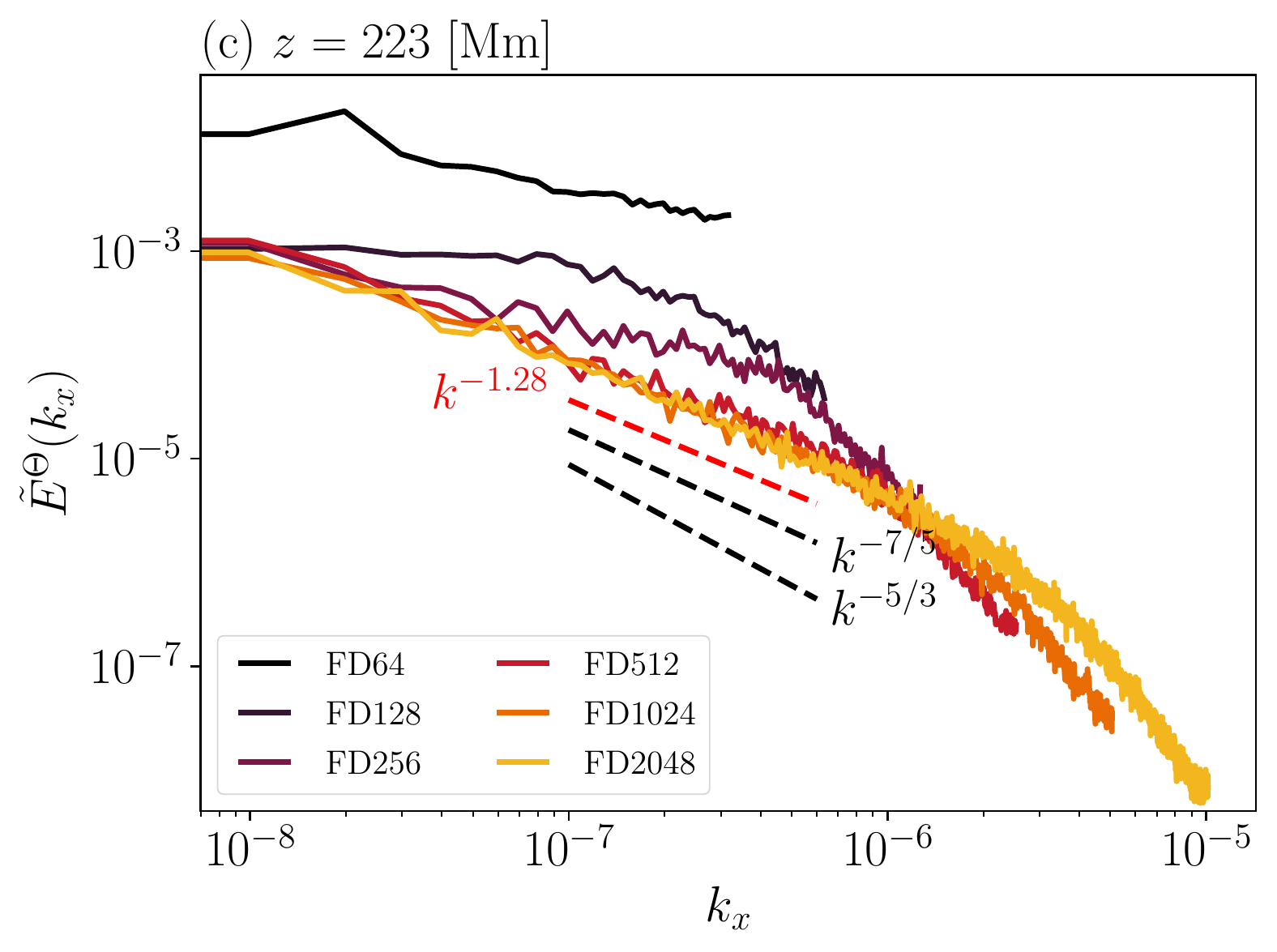} 
    \includegraphics[width=0.87\columnwidth]{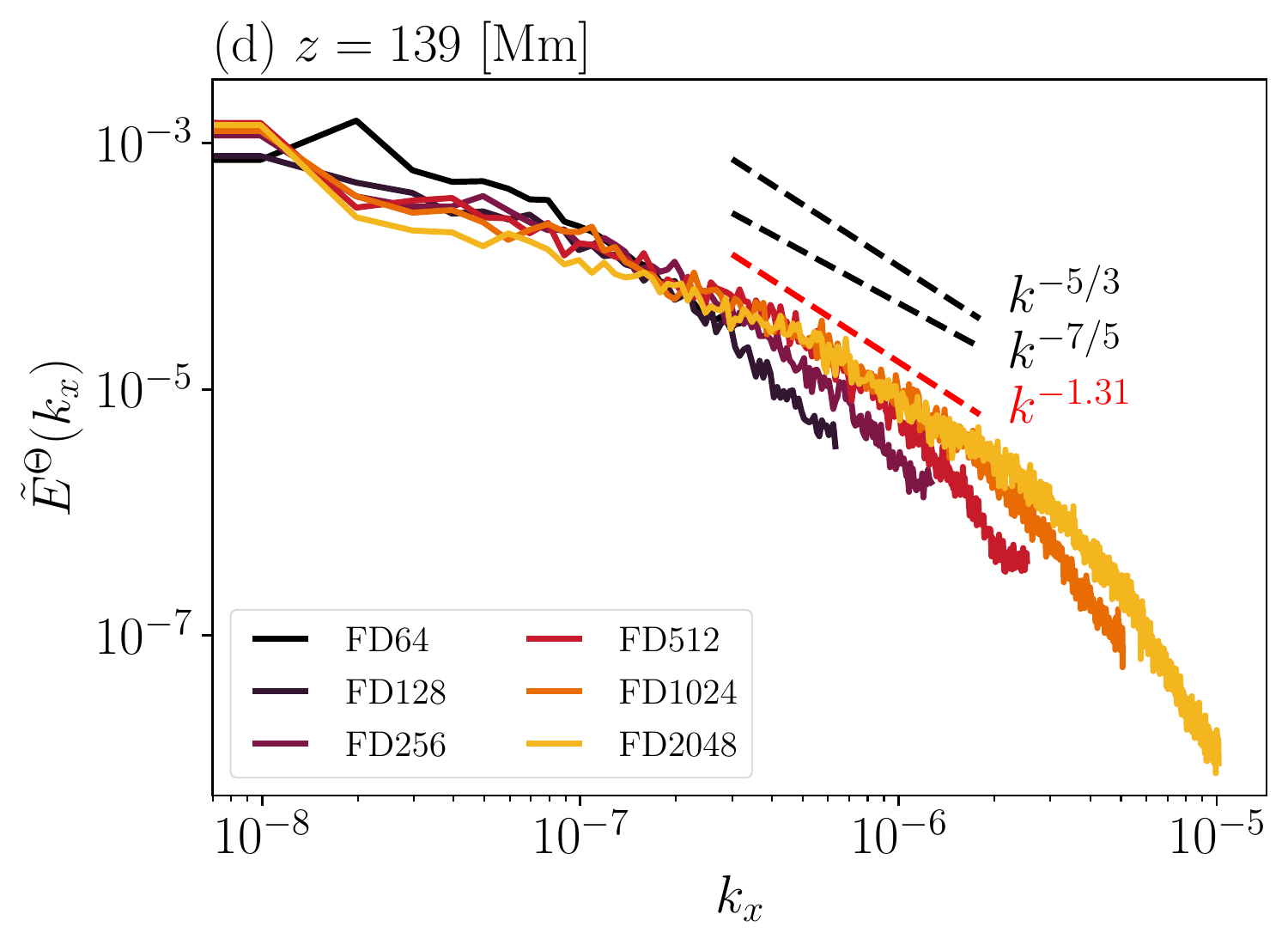} 
    \caption{Power spectra of the kinetic energy, upper panels,  and the variance of potential 
    temperature, bottom panels, for different numerical resolutions. Panels (a) and (c) correspond
    to the upper part of the domain, $z=223$ Mm; and panels (b) and (d) to the middle of
    the convection zone, $z=139$ Mm.
    The black dashed lines are guides to the KO and BO scaling laws, and the red dashed lines
    correspond to the adjusted scaling for simulation FC1024 in the inertial range.}%
    \label{spectra_nc}%
\end{figure*}

A comparison of the kinetic power spectrum of simulations with different resolutions  is
presented in Fig.~\ref{spectra_nc} for $z=223$ Mm (a) and $z=139$ Mm (b) as a function of 
the wavenumber; panels (c) 
and (d) display the power spectrum of the variance of potential temperature, 
$\tilde{E}^{\Theta}(k,z)$ for the same heights. 

In 3D isotropic turbulence, $\tilde{E}^{K}(k)$ and $\tilde{E}^{\Theta^{\prime}}(k)$ 
scale according to the Kolmogorov law, as $k^{-5/3}$ \citep[][hereafter KO scaling]{k41,obukhov59}.
In the presence of the buoyancy force, but still for isotropic motions, the
scaling should be $\tilde{E}^{K}(k)\sim k^{-11/5}$ and 
$\tilde{E}^{\Theta^{\prime}}(k) \sim k^{-7/5}$ \citep[][hereafter the BO 
scaling]{bolgiano59,obukhov59}.
Our simulations are in 2D and the 
motions are clearly anisotropic as it can be observed in Fig.~\ref{snapshots_w}. 
Therefore, none of the scaling laws above should be applicable. Nonetheless, for 
the sake of comparison the black dashed lines in 
Fig.~\ref{spectra_nc} compare these scaling laws with the scaling found
for the inertial range in our simulation with $N=1024$ (in red). At both heights the
kinetic and thermal spectra are closer to the BO than to the KO scaling laws.

It is noticeable that the kinetic spectra at the upper levels of the convection zone, 
panel (a) have inertial
range starting at almost the largest scales, whereas in the middle of the convection zone,
panel (b), the inertial range starts at $k\sim10^{-7}$ m$^{-1}$.   
As expected, the inertial range extends over more wavenumbers as the resolution increases, 
reaching the dissipative Kolmogorov scale at the smallest resolved scales.
The scaling law of the energy in the inertial range changes from one depth to the other.  
This change might be a manifestation of differences in the anisotropy of convective motions at 
different depths.  
Except for the simulation FD64, there is a good agreement between results of 
different simulations in the kinetic power 
in most of the inertial range at $z=223$ Mm and at the large scales, $k \lesssim 10^{-7}$ m$^{-1}$, 
for $z=139$ Mm.
Our results are in contrast to the DNS simulations performed by 
\cite{2016ApJ...818...32F}, where the large scales lose energy when the dissipative 
coefficients are diminished and smaller scales are resolved. However,
there is an agreement between our findings, i.e., convergence of the spectra
for the large scales, and the results of the LES simulations of \citep{sullivan+11}. 
This suggests that SGS or implicit SGS methods properly
capture the inverse energy cascade from the smaller to the larger scales.

The profiles of $\tilde{E}^{\Theta^{\prime}}(k)$ in Fig.~\ref{spectra_nc}(c) present
evidence of the lack of resolution discussed above in simulations FC64-FC256 at the upper layers 
of the domain where the density scale height is small. 
It is manifested in excess of energy at all scales for FC64 and at intermediate
scales for FC128 and FC256. On the other hand, simulations FC512-FC2048 show convergence in the 
inertial range of the power spectra.  
In the middle of the convection zone, where the 
density scale height is larger, most of the simulations have similar power spectra, differing
only in the extent of the inertial range (see panel d).

\subsection{Numerical viscosity} 
\label{sec:sgs}

We estimate the effective viscosity, $\nueff$, for the simulations 
with all the tested resolutions. There are several techniques to conduct this calculation
for ILES simulations, most of them performed with the physical quantities 
represented in the Fourier space \citep[e.g.,][]{zhou2014estimating}, but some 
methods also perform this computation in both physical and spectral
spaces \citep{SCHRANNER201584}.
\cite{domaradzki2003effective} presented a method to compute the numerical
viscosity for an EULAG simulation of decaying turbulence. 
Here we adapt the method implemented by \cite{domaradzki2003effective} and 
developed later for convection simulations in spherical coordinates by
\cite{2016AdSpR..58.1538S} to our Cartesian 2D simulations. Besides providing 
the profile of $\nueff$ as a function of  the wave numbers for each resolution, i.e.,  
the amount of viscosity for different scales, the method allows 
a clear comparison of this quantity as the numerical resolution varies.

To estimate the effective viscosity we  
Fourier transform all terms in Eq.~\eqref{eq_momentum_nc},  and take the
dot product on both sides with $\rho_r \textbf{u}^{\ast}_{k}$, where 
$\textbf{u}_{k} = \textbf{u}(t,k,z)$. Then, we average the resulting equation with 
respect to $t$ and $z$, and add the contribution of the effective 
viscosity to obtain the following equation,
\begin{equation}
    \frac{\partial\varepsilon_{k}}{\partial t} = 
    \mathcal{A}_{k} +  \mathcal{P}_{k} + \mathcal{G}_{k} + \mathcal{D}_{k}(\nueff)
    \label{eq_nu_1}
\end{equation}

\noindent where the kinetic density energy term is given by

\begin{equation}
    \varepsilon_{k} = \frac{1}{2}\langle\rho\textbf{u}_{k}\cdot\textbf{u}_{k}^{\ast}\rangle,
\end{equation}

\noindent while the terms corresponding to advection, pressure, gravitational 
potential energy and the effective viscous dissipation rates are, respectively, 
given by
\begin{equation}
    \mathcal{A}_{k} = 
    -\langle\rho_r \left[(\textbf{u}\cdot\nabla)\textbf{u}\right]_{k}\cdot\textbf{u}_{k}^{\ast}\rangle,
\end{equation}

\begin{equation}
   \mathcal{P}_{k} = -\langle\rho_r (\nabla\pi^{\prime})_{k}\cdot\textbf{u}_{k}^{\ast}\rangle,
\end{equation}

\begin{equation}
    \mathcal{G}_{k} = -\left\langle\rho_r \textbf{g}\left(\frac{\Theta^{\prime}}{\Theta_{r}}\right)_{k}\cdot\textbf{u}_{k}^{\ast}\right\rangle,
\end{equation}

\noindent and

\begin{equation}
    \mathcal{D}_{k}(\nu) = \left\langle \left\{\frac{\partial}{\partial x_{j}} \left[2\rho_r\nu \left(e  - \frac{1}{3} \delta_{ij}\nabla\cdot\textbf{u}\right) \right]\hat{\textbf{e}}_i\right\}_{k}\cdot\textbf{u}^{\ast}_{k}\right\rangle,
    \label{eq_nu_d}
\end{equation}

\noindent where $\langle\cdot\rangle$ means the average over $z$ and time.

If the averages are taken during the statistically steady state and over a long period 
of time, the left-hand side of Equation~\eqref{eq_nu_1} is approximately zero. 
Thus, dividing the remaining 
equation  by 
$-\mathcal{D}_{k}(\nu=1)$, reorganizing,  and overlooking the fact
that $\nueff$ might be dependent on the space coordinates, we obtain an estimate of 
the effective viscosity of the system, 
\begin{equation}
    \frac{\mathcal{A}_{k} +  \mathcal{P}_{k} + \mathcal{G}_{k}}{-\mathcal{D}_{k}(\nu=1)} =  \frac{-\mathcal{D}_{k}(\nu=\nueff)}{-\mathcal{D}_{k}(\nu=1)} =
    \nueff(k).
    \label{eq_nu_d_3}
\end{equation}

The reader should be aware of the several approximations made to
compute what we call numerical or effective viscosity throughout this paper. In ILES 
the SGS viscosity is nonlinear, intermittent in space and time. On the other hand, 
our estimate for this contribution, $\nueff$, is considered constant with 
the depth and results from spatial and temporal 
averages. Furthermore, in the MPDATA algorithm implemented in the EULAG-MHD code, 
besides the dissipative terms,  there are dispersive SGS contributions, effectively 
responsible for inverse cascades \citep{margolin+02}. These terms are evidently not 
captured by Eq.~(\ref{eq_nu_d}).
Therefore, $\nueff$ is an integral measure of the SGS viscosity which also embodies
other sub-grid scale contributions implicit in the numerical technique.

\begin{figure}%
    \centering
    \includegraphics[width=0.85\columnwidth]{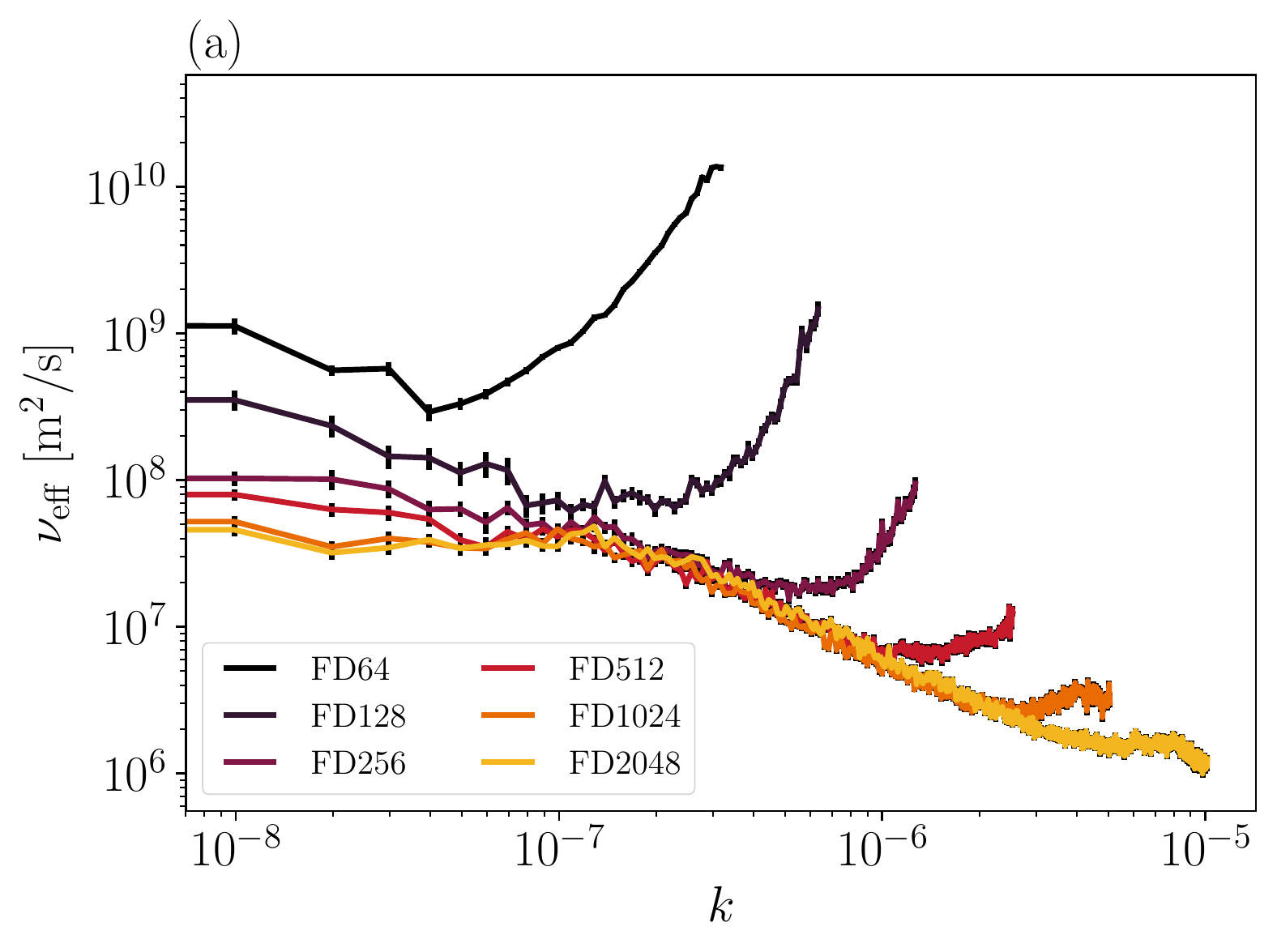}\\ 
    \includegraphics[width=0.85\columnwidth]{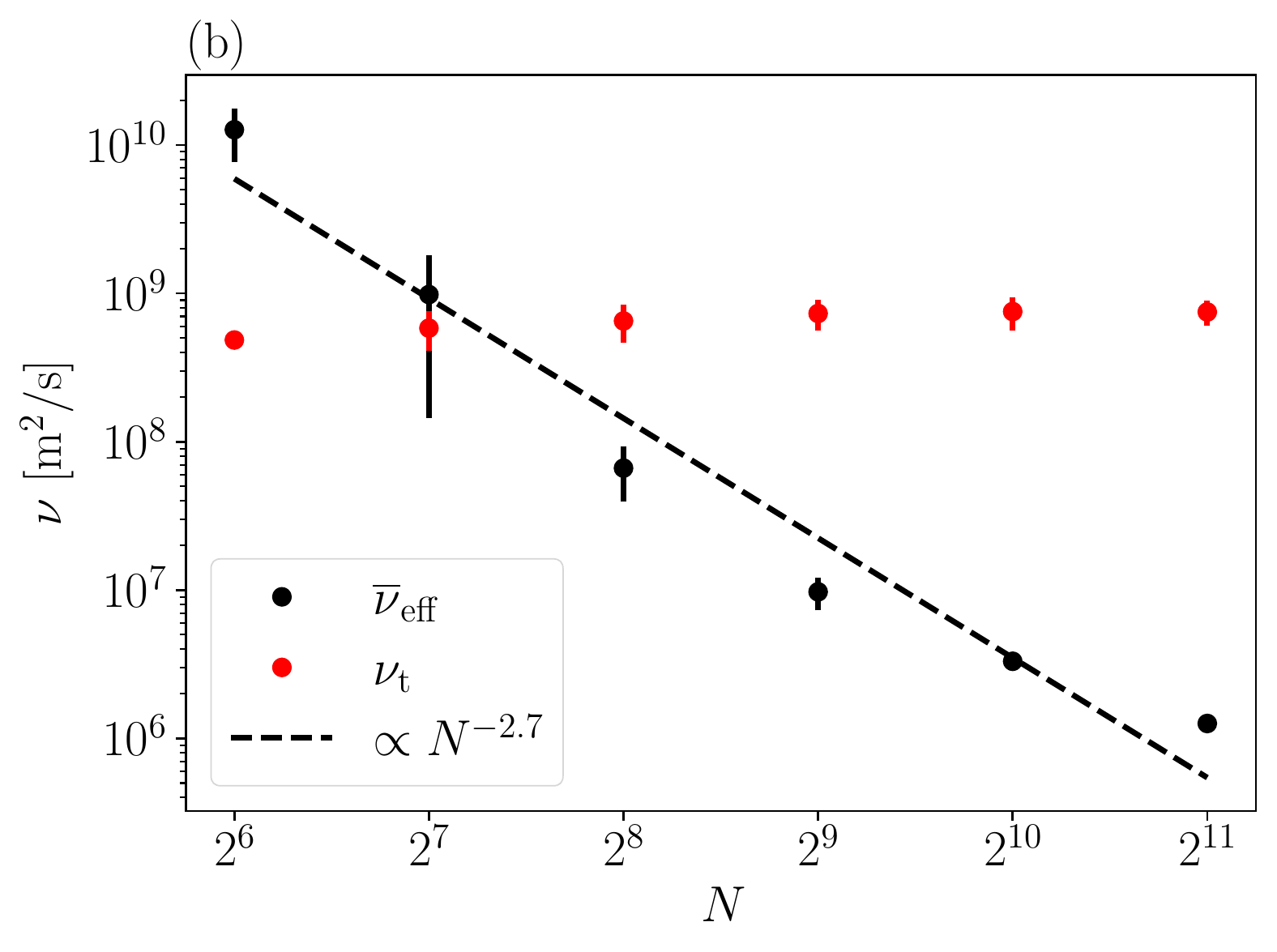}
    \caption{(a) Effective numerical viscosity as a function of the wave number, $k$, for  
	simulations with different resolutions; (b) effective numerical viscosity (black points) 
	of the smallest
	resolved scales and turbulent viscosity (red) versus the resolution, $N$.  
	Upper and bottom panels show vertical error bars 
	multiplied by a factors of 2 and 10, respectively, to make them noticeable. The black dashed 
	line shows a fitted 
	power law as indicated in the legend. }%
\label{visc_comp}%
\end{figure}

In Fig.~\ref{visc_comp}(a) we present $\nueff(k)$ for the simulations with 
different resolutions.  For the large scales, $k < 10^{-7}$ m$^{-1}$,
$\nueff$ decreases from $10^9$ m$^2$/s to $6\times10^7$ m$^2$/s. The difference
in $\nueff$ between the different simulations decreases as the resolution increases.
For simulations FD64 to FD256, the profiles of $\nueff$ have a minimum for the
intermediate scales and an increase for the smallest scales. Qualitatively,
these profiles resemble the results of \cite{domaradzki2003effective}. 
The profiles obtained by \cite{2016AdSpR..58.1538S} for a global simulation with 
$51\times64\times128$ grid points also show increasing 
effective viscosity for large wave-numbers. However, in their more sophisticated three 
dimensional procedure they obtain large errors for the low wave-numbers and disconsider 
the results for these scales. A comparison of the values of $\nueff$ averaged over 
large wavenumbers shows quantitative agreement between the results of 
\cite{2016AdSpR..58.1538S, Strugarek+18} and our simulations FD128 and FD256.
The results of \cite{Strugarek+18} show that the effective viscosity depends on 
the strenght of convection, the density constrast and/or rotation. Yet, this dependency
is weak when compared to the changes found here for different resolutions.

For the cases FD256 to FD2048, there is an intermediate range of $k$, for which 
$\nueff$ is the same for all resolutions.  For the smallest scales, it 
has decreasing values with the increase of the numerical resolution. 
For FD512-FD2048, the profile of $\nueff$ does not increase but remains roughly
constant.  The Fig~\ref{visc_comp}(a) shows error bars for this estimation 
multiplied by 2 to make them distinguishable. The error is computed as $\sigma/\sqrt{n}$, where 
$\sigma$ is the standard deviation of the temporal average and $n$ is the 
number of temporal samples used in the computation corresponding to the last 15 yr
of the simulations. Thus, $n$ is of the order of $50$, for the considered sampling times.

The magnitude of the effective viscosity averaged over the largest wave numbers 
resolved for each simulation, i.e., over the Kolmogorov scales, is  
presented in Table~\ref{table_results} and depicted in Fig.~\ref{visc_comp}(b) 
as a function of $N$. It decreases as the resolution increases following a 
power law, $\overline{\nu}_{\rm eff} \propto N^{\alpha}$,  with exponent $\alpha = -2.7$. However, 
the power is higher for the low-resolution cases and smaller for
simulations with $N>512$.  With the value of $\overline{\nu}_{\rm eff}$ and $\urms$ 
(see Table ~\ref{table_results}) 
we can compute an effective Reynolds number, $\Re_{\rm eff}=\urms L/\overline{\nu}_{\rm eff} $ (with 
$L=0.3\Rs$, i.e., the size of the convection zone), reached by the simulations.  Its values 
go  from $\sim1$ to $\sim 7\times10^3$, and its variation as a function of $N$, presented in
Fig.~\ref{reynolds}, follows a power law $\Re_{\rm eff} \propto N^{2.7}$. 

\vfill\eject

\subsection{Turbulent Viscosity}
\label{sec:turb_visc}

Even though the effective viscosity decreases with the resolution following 
a power law, Figs.~\ref{urms_nc} and \ref{spectra_nc} evidence 
convergence of the simulation results. This implies that, despite the smallest
values of $\nueff$ in the high-resolution cases, the dynamics of the system 
is governed by an enhanced dissipation, likely provided by turbulence, which
efficiently transports momentum and heat.
The turbulent viscosity,  $\nut$,   depends on the time 
and spatial scales of the most energetic eddies, which are model-dependent
quantities. They are 
sensitive to variations in the domain's aspect ratio, the ambient state and the 
time scale of the thermal relaxation \citep{cossette+16}.
In this section, we determine $\nut$ to verify if it also depends on the 
resolution and how its values compare with those of $\nueff$. 

To do so, we calculate first the turbulent correlation length of the convective 
motions using \citep{2000tufl.book.....P},
\begin{equation}
    \ell = \frac{\int \tilde{E}(k)/k dk}{\int \tilde{E}(k) dk}.
    \label{eq_corr_len}
\end{equation}

Equation \eqref{eq_corr_len} can be seen as a weighted average of the inverse of the 
wavenumbers, where the weights
are given by the kinetic energy. Therefore it provides the typical length of the most 
energetic convective eddies. Their values vary between $\sim 40$ and $\sim 50$ Mm as 
presented in Table~\ref{table_results}.

With the correlation length, $\ell$,  we estimate the turbulent viscosity as 
\citep[see, e.g.,][]{kitchatinov+94}
\begin{equation}
    \nu_{\rm t} = \frac{1}{3}\ell u_{rms}.
    \label{eq_turb_visc}
\end{equation}
The values of $\nut$ as a function of $N$ are presented as red points in Fig.~\ref{visc_comp}(b).
They slightly rise as the resolution increases from $\sim 5\times 10^8$ to $\sim 8\times 10^8$ m$^2$/s.
These values have the same order of magnitude of recent estimations of the turbulent magnetic 
diffusivity due to granulation and supergranulation in the solar surface \citep{skokic+19} which
suggest that our model parameters are in an appropriate regime.

\begin{figure}[htb]%
    \centering
    \includegraphics[width=0.9\columnwidth]{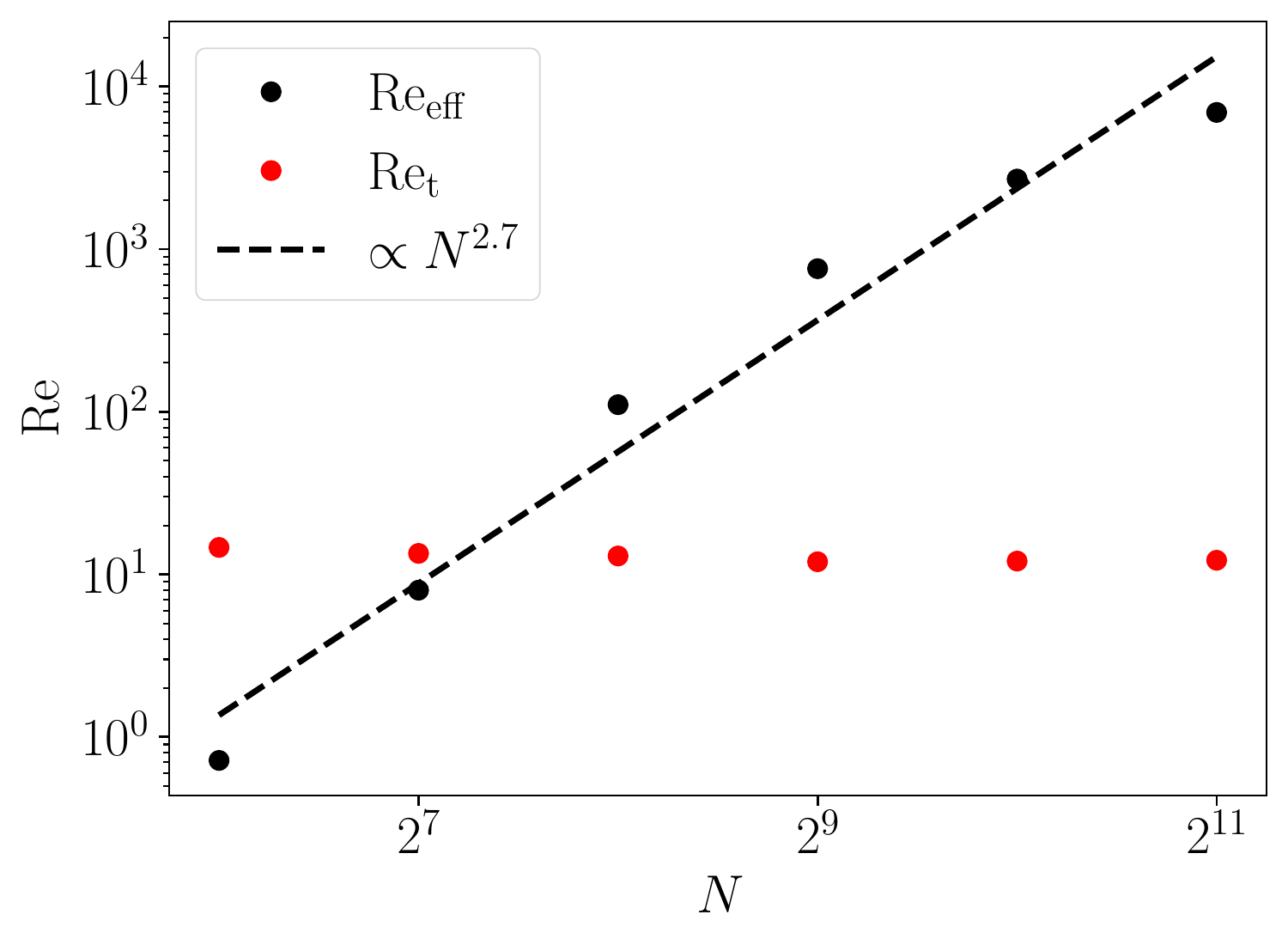}\\
    \caption{Effective Reynolds number (black points) and turbulent Reynolds (red) number 
	as a function of the resolution. The black dashed line shows a fitted power law as 
	indicated in the legend.}
    \label{reynolds}%
\end{figure}

The variance of $\nut$ for different resolutions is rather small when compared to the changes of 
$\overline{\nu}_{\rm eff}$ which span about 4 orders of magnitude. As a consequence, the 
Reynolds number computed 
from the turbulent viscosity remains roughly constant with 
values $\sim 13$ while the effective Reynolds number increases 4 orders of magnitude,
see Fig.~\ref{reynolds}.
From Fig.~\ref{visc_comp}(b), it can be seen that only the 
simulation FD64 has effective viscosity considerably larger than the turbulent one. 
For all the other cases, $\nut>\overline{\nu}_{\rm eff}$ implies that the system is 
governed by the large scales. 
The relevant question is, why do these scales, especially in the bulk of the convection zone,
have similar spatio-temporal correlations regardless of the resolution? The results presented in
sections \ref{sec:phys}-\ref{sec:sgs} indicate convergence for
simulations FD512-FD2048 with well-resolved dynamics. For simulations  FD128 and FD256, the
agreement is not perfect, but the large scales behave alike.  This might be
a consequence of the implicit SGS contribution towards the dynamically dominant scales.  
The different profiles of $\nueff(k)$ between these two sets of simulations provide 
some support to this hypothesis; note the increasing values of $\nueff$ for the smallest 
scales in models with $N=128,256$.

\subsection{Internal gravity waves and mean flows in the stable layer}
\label{sec:qbo}

\begin{figure}%
    \centering
    \includegraphics[width=0.96\columnwidth]{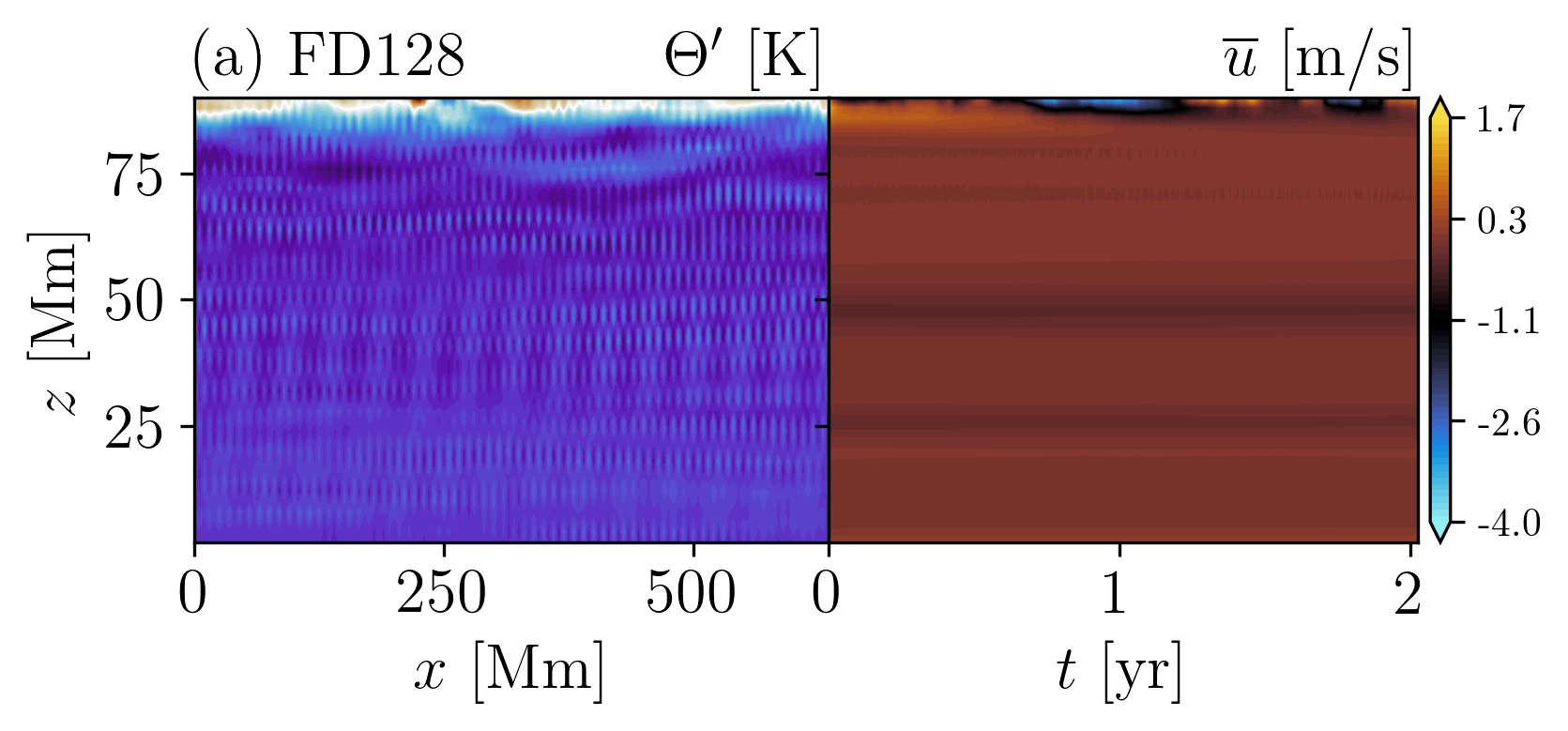} \\
    \includegraphics[width=0.96\columnwidth]{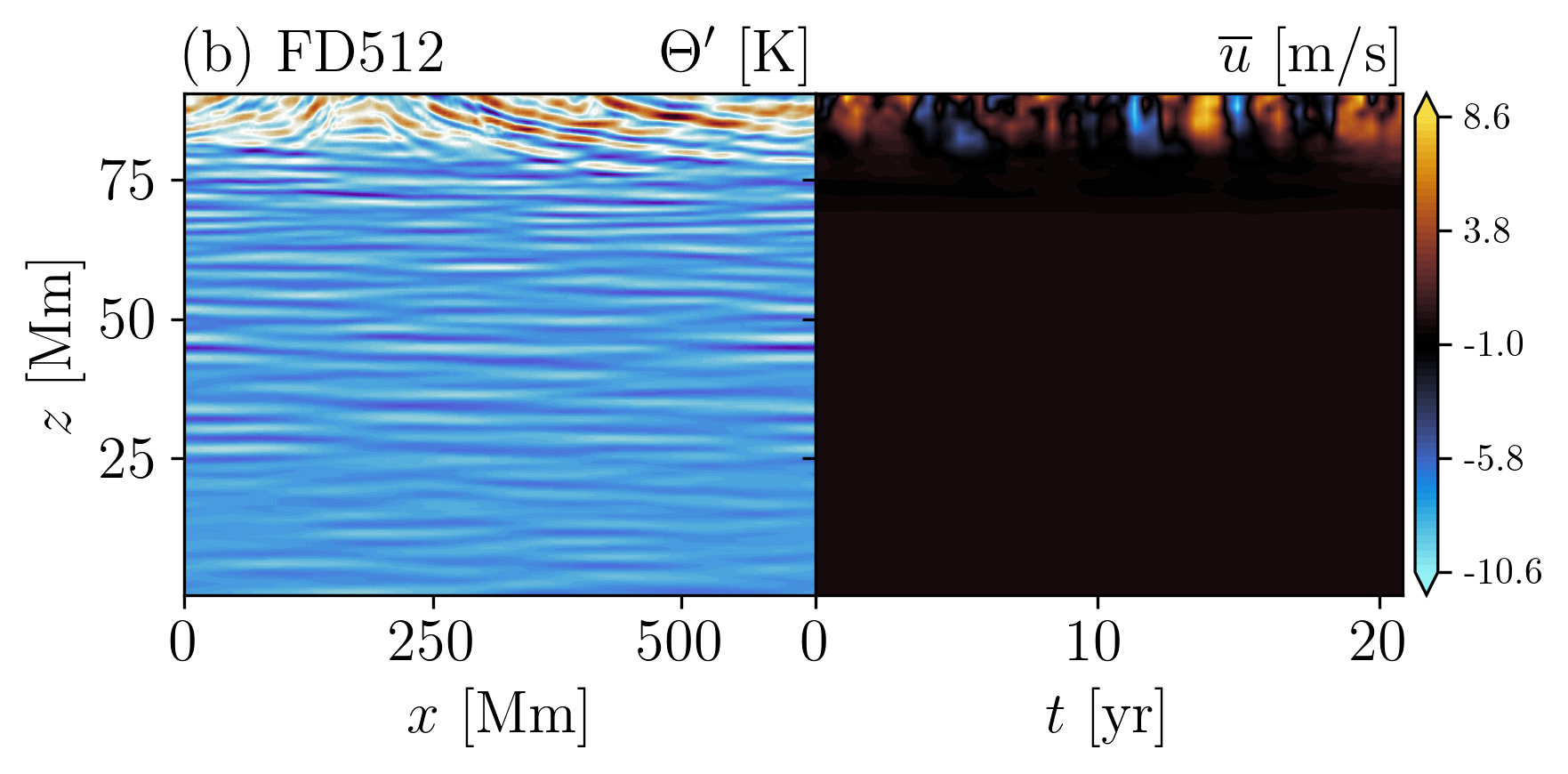} \\
    \includegraphics[width=0.96\columnwidth]{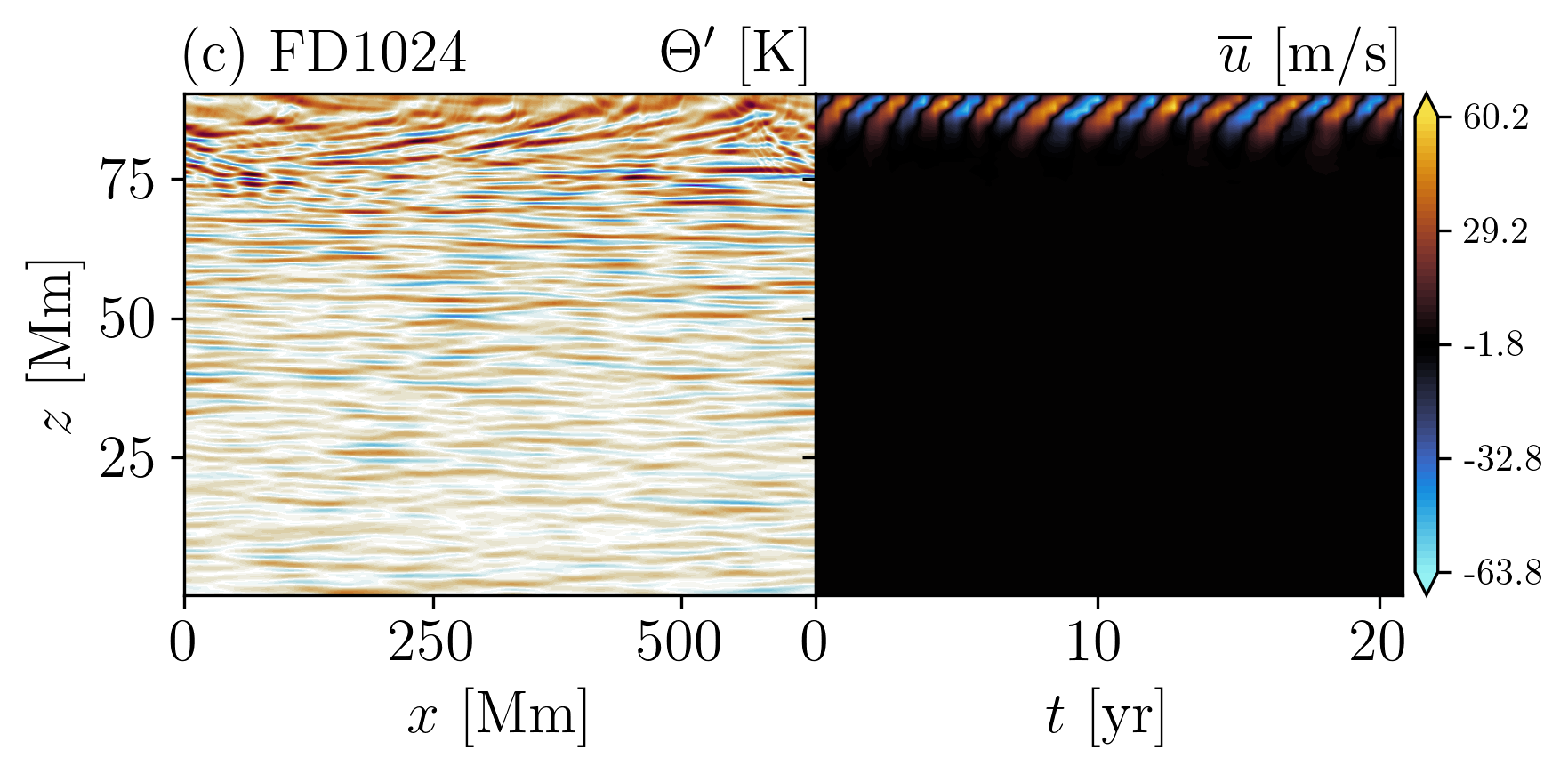} \\ 
    \includegraphics[width=1.0\columnwidth]{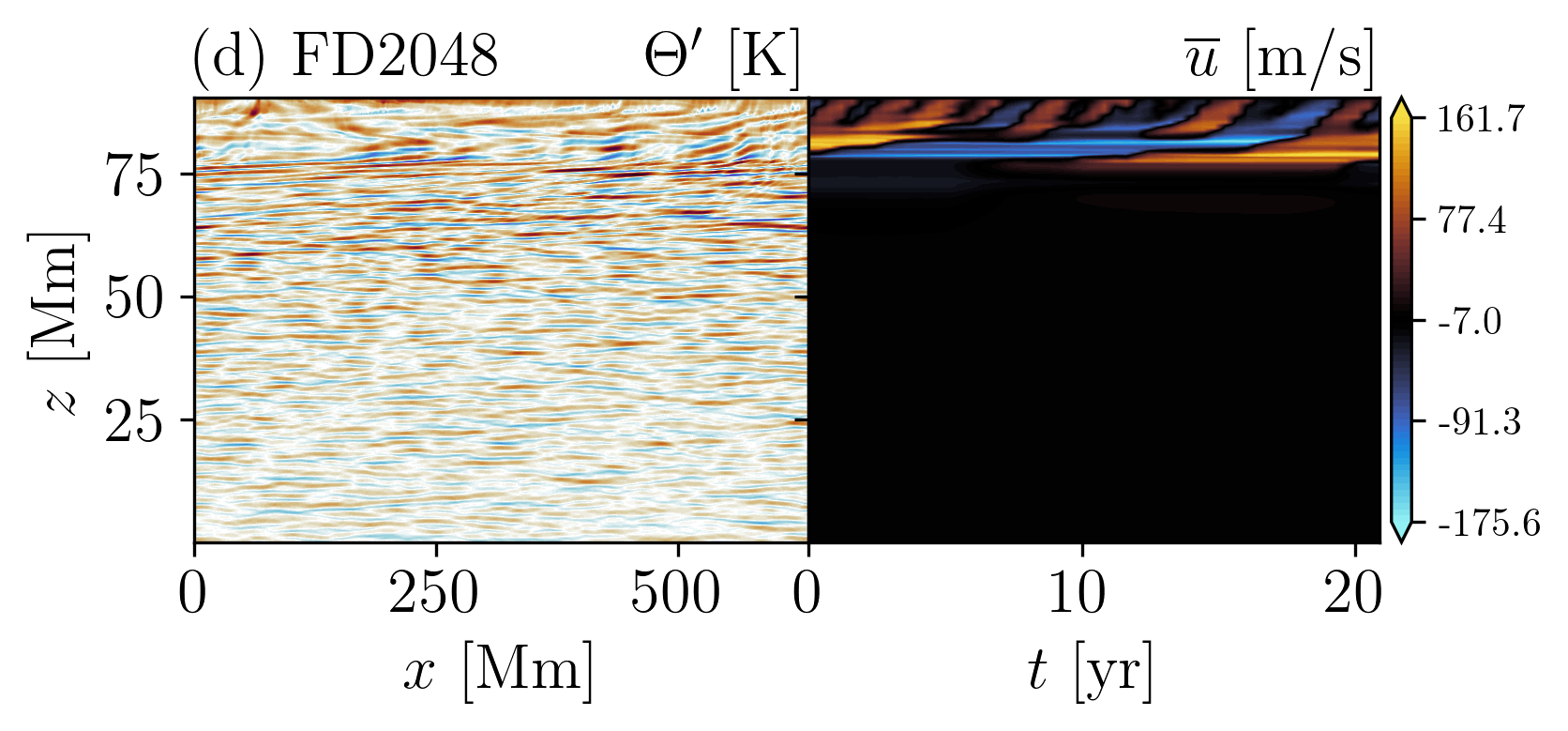}
	\caption{ Left panels: snapshots of $\Theta^{\prime}$  for simulations  (a) FD128, 
	(b) FD512, (c) FD1024, and (d) FD2048. The aspect ratio of this panel is modified for
	visualization purposes. Right panels: Propagation in the plane ($t,z$) of 
	the velocity component, $u$,  averaged over the horizontal direction, $x$. The 
	time corresponds
	to the last 20 years of evolution during steady state.}%
    \label{gwqbo}%
\end{figure}

As a consequence of convective overshooting, internal gravity waves are excited and
propagate in the stable layer.  Interestingly, together with these waves a mean horizontal
motion develops in the stable layer. In our high-resolution experiments this motion
reverses sign periodically as it can been seen in Fig.~\ref{gwqbo}. 
The left panels show snapshots of $\Theta^{\prime}$ in the stable layer. 
The gravity waves are unresolved in the simulation FD128, barely captured in 
FD512, and  resolved in FD1024 and FD2048. The right panels show
the temporal evolution of  $\overline{u}$, where the overline represents average 
over the horizontal direction, during the last 20 years of the simulated time. For the case
FD128 there is an unorganized, low amplitude pattern at the very upper fraction
of the stable interior.  For FD256 the mean flow is more evident showing non-periodic 
reversals. For FD1024 a well organized pattern of mean flow reversing sign every
$\sim 1$ yr emerges. The amplitude of this flow reaches $60$ m/s which is of the same order
of the horizontal motions in the convection zone.  Finally, because of a small viscosity, 
for FD2048 these motions have large amplitudes, change direction in a random form and
persist for longer time at $z \sim 75$ Mm  forming two or three layers of mean-flow, 
$\overline{u}$,  at the same time. Similar results were discussed for small viscosity
simulations in \cite{wedi+06}. These motions are reminiscent of the
Earth quasi-biennial oscillation \citep{baldwin01} which are a 
consequence of interacting gravity waves and depend on the kinematic viscosity of
the medium \citep{lindzen+68,holton+72,plumb+78,wedi+06,kim+01}.
If existing inside the solar radiative zone, they may be relevant for the 
transfer of angular momentum and may interact with large scale motions in the convection zone, 
likely contributing to the formation of torsional patterns. Since the goal of this work is 
exploring the role of numerical resolution, we postpone a 
more detailed physical discussion on the physics of GW and these mean-flows 
for future works.  

\section{Conclusions}
\label{sec:conclusions}

Direct numerical simulations of natural systems at high Reynolds numbers are unattainable 
for present supercomputers.  Fortunately, the LES and 
ILES methods provide capabilities of reproducing 
turbulent flows. However, for systems with scarce observational constraints, such as 
stellar convection zones, determining whether LES or ILES capture the system’s physics 
is a challenging task.

The aim of this paper was to
address this problem by exploring the numerical convergence in ILES 
simulations of turbulent convection in two dimensions.  The model is constructed over
the anelastic set of equations solved by the EULAG-MHD code. 
It considers an atmosphere with characteristics of the solar interior, including
a fraction of the radiative zone and the convection zone.  
The simulations were performed with resolution increasing from $64^2$ up to $2048^2$ grid points.
The results present the values of the effective and turbulent viscosities and 
other integral characteristics of the numerical solutions. Another goal was to observe how the large 
scales behave when  interacting with the progressively smaller resolved scales.

Spatial and temporal averages demonstrate that quantities as the rms velocities 
have similar vertical profiles for resolutions even as coarse  as $N\sim128$. 
We noticed that the structure of the flow, characterized by narrow downdrafts
and broad upflows crossing the entire convective layer of the model, is conserved 
even when these motions interact with the smallest structures (Figs.~\ref{snapshots_w} and 
\ref{snapshots_the_nc}). This result stands in contrast to the results of 
2D ILES of \cite{porter+94} where the progressive development of small scales
leads to the destruction of the large structures. This difference
might arise from the differences between PPM and MPDATA ILES formulations, 
but mostly  from the compressible character of their simulations which imposes
structural changes in the temperature and density profiles as the effective viscosity
decreases. These changes result in different convective models
for each resolution. 

As for the spectral behavior, we notice that for resolutions with $N\gtrsim 128$, 
the kinetic energy spectra are similar at the middle and the top of the convection
zone, whereas the length of the inertial range increases with $N$ reaching more than 
two decades for our highest resolution model. As for the turbulent spectrum of the 
variance of $\Theta'$, in the top, it reflects the lack of resolution of
the FD64-FD256 simulations due to the small density scale height. Convergence is 
observed, however, in simulations with  $N\gtrsim 512$. In the middle of the convection
zone, simulations with $N=128$ or more grid points agree in the injection and inertial
ranges.

A deeper analysis of the energy balance in the Fourier space allows us to 
estimate the effective viscosity of the simulations as a function of the 
wavenumber.  For the large scales, the profiles of $\nueff$ in all the simulations 
are slightly decreasing with $k$, and with magnitudes progressively decreasing with
$N$.  For $N \ge 256$ there is a range of intermediate scales where the values and 
profiles of $\nueff$ closely match. These profiles separate for the smallest resolved scales. 
For the smallest scales the profiles of $\nueff$ increase with $k$ 
for resolutions with $N \le 256$ but remain roughly constant for $N \ge 512$. Our interpretation 
of this change has to do with the simplified assumptions made for the residual of the 
balanced Eq.~(\ref{eq_momentum_nc}) in the steady state. We assumed that
the residual corresponds only to an effective viscosity depending on the 
scale but constant in time. 
However, in MPDATA the numerical formulation contains dissipative
as well as dispersive terms, both intermittent in time and space. 
The contribution of the dispersive terms is mediating  
the transfer of energy between scales. Thus, the increasing of $\nueff$ 
for the smaller scales in the low-resolution cases can be related to the enhanced 
sub-grid scale contribution of the numerical algorithm. Yet, the small-scale 
contribution is thoroughly resolved at higher resolutions. Therefore, the profile of $\nueff$ 
would incorporate only viscous dissipation.

Averaging $\nueff(k)$ over the dissipative, Kolmogorov, scales 
shows a power law relation between the effective viscosity and 
the resolution, $\overline{\nu}_{\rm eff} \propto N^{-2.7}$. Note, however, that the curve could
be fitted by 2 different power laws for $N \le 256$ and $N \ge 512$, which
supports our conclusion above on the SGS contributions.  With the obtained values of 
$\overline{\nu}_{\rm eff}$ the
effective Reynolds numbers the simulations span between $\sim1$ and $\sim 10^{4}$.
On the other hand, the values of the turbulent eddy viscosity, $\nut$, are of the order of 
$10^8$ m$^2$/s and are barely dependent of the grid size.  Only the simulation FD64 has average
effective viscosity larger than this value.  Thus, in spite that small-scale structures
are sharply resolved in high-resolution cases, the dynamics of the system is determined by the 
turbulence which has an eddy Reynolds number of the order of $10$.  We argue that 
the diagnostics observed for simulations FD512-FD2048 indicated convergent well-resolved
turbulence. The fact that the properties of the large scales in simulations FD128 and 
FD256 behave alike to the better resolved ones evidences the SGS contribution of
the advection solver MPDATA. The convergent LES simulations of \cite{porter+00,sullivan+11} 
obtained results in agreement to the ones outlined here.  Alternatively, the DNS simulations 
of \cite{2016ApJ...818...32F} indicate that decreasing the dissipation coefficients
diminishes the energy of the large scales in benefit of the smallest ones. 

Despite the simulations presented here being 2D, the results of this research support 
the idea that ILES are efficient in capturing the dynamics of turbulent systems.
A note of caution is due here for regions where the thermal stratification enforces
the formation of small structures not well resolved by the grid as in cases
FD64 and FD128.  Having a non-homogeneous grid could be appropriate for these
situations.   Nevertheless, if the model encompasses both stable and unstable 
layers, as in our case, high resolution is necessary to 
capture the dynamics in the stable layer dominated by gravity waves.
This is demonstrated in \S~\ref{sec:qbo} where mean oscillatory motions emerge only for 
sufficiently low effective viscosities. 
A relevant question raised by this study is how much the interaction between these two layers modifies
the dynamics of the turbulent convection.  Simulations of convection in 3D and/or in 
different geometries that may help answering this question will be explored in future work.

\acknowledgments
We thank the anonymous referee for the constructive comments and
suggestions that have improved the quality of the paper.
HN thanks CNPq for financial support. 
This work was partly funded by NASA grants NNX14AB70G,  80NSSC20K0602, and 
80NSSC20K1320.  NCAR is sponsored by the National Science Foundation.
The simulations were performed in the NASA supercomputer Pleiades.


\bibliographystyle{aasjournal}
\bibliography{bib}

\end{document}